\documentclass[aps,pra,reprint,superscriptaddress,floatfix,preprintnumbers,showkeys]{revtex4-1}

\usepackage{graphicx}
\usepackage{color}
\usepackage{amssymb}
\usepackage{mathtools}
\usepackage{bm}
\usepackage[american]{babel}
\usepackage{hyperref}
\usepackage{braket}
\usepackage{siunitx}
\usepackage[T1]{fontenc}
\usepackage{txfonts}

\DeclareGraphicsExtensions{.pdf}
\hypersetup{colorlinks=true,urlcolor=blue,breaklinks=true,citecolor=blue,linkcolor=blue,pdfstartview=FitH,pdfpagemode=UseNone}
\mathtoolsset{showonlyrefs=false}

\newcommand{\intmeasure}{r\,\mathrm{d} r \,\mathrm{d} \phi } 
\newcommand{\ud}{\mathrm{d}}
\newcommand{\abs}[1]{\lvert #1 \rvert} 
\newcommand{\componentnumber}{K} 
\newcommand{\osclength}{a_{\mathrm{osc}}} 
\newcommand{\bohrradius}{a_{{\mathrm{B}}}} 
\newcommand{\amplitude}{f}
\newcommand{\windingnumber}{\kappa}
\newcommand{\qpindexa}{q} 
\newcommand{\qpwinding}{l} 
\newcommand{\maxqpwinding}{l_{\mathrm{dom}}} 
\newcommand{\bogoper}{\mathcal{B}} 
\newcommand{\axisymoperator}{\mathcal{D}}
\newcommand{\polarization}{P}
\newcommand{\interparam}{\Gamma}
\newcommand{\geff}{g_{\mathrm{eff}}}
\newcommand{\intrag}[1]{g_{{#1}{#1}}} 
\newcommand{\intraa}[1]{a_{{#1}{#1}}} 
\newcommand{\perturbation}{\chi}
\newcommand{\tcv}[2]{({#1},{#2})} 
\newcommand{\thcv}[3]{({#1},{#2},{#3})} 
\newcommand{\pointcoord}[2]{\left(#1,#2\right)} 
\newcommand{\openinterval}[2]{\mkern+1mu]\mkern-0.5mu{#1},{#2}\mkern-0.5mu[\mkern+1mu} 
\newcommand{\ntot}{n_{\mathrm{tot}}} 
\newcommand{\norm}[1]{\lVert #1 \rVert} 
\newcommand{\parameterset}[1]{S_{\mkern-2mu{#1}}}
\newcommand{\replacedwith}{\rightarrow} 
\newcommand{\myDots}{\ifmmode\mathinner{\ldotp\kern-0.2em\ldotp\kern-0.2em\ldotp}\else.\kern-0.13em.\kern-0.13em.\fi}

\newcommand{\dynstabledescription}{All states classified as dynamically stable in panel~(b) are energetically unstable. }
\newcommand{\enestabledescription}{The energetic stability indicated by the star symbols is local but not global. }

\newcommand{\stabilitydiagramcaption}[4]{(a) Maximum imaginary part of the excitation frequencies $\{\omega_\qpindexa^{(\qpwinding\mkern+1mu)}\}$ of the stationary $\tcv{#1}{#2}$ vortex as a function of the relative intercomponent interaction strength $\interparam\coloneqq g_{12}/\sqrt{\intrag{1}\intrag{2}}$ and the particle-number polarization $\polarization\coloneqq(N_1-N_2)/(N_1+N_2)$. The colorbar scale is in units of $\omega_1$. (b)~The angular-momentum quantum number $l$ that provides the dominant dynamical instability. {#3}{#4}The values of the other parameters of the model are taken from set~$\parameterset{1}$, as detailed in the caption of Fig.~\ref{fig:maxim-2-0}.}

\newcommand{\timeseriescaption}[9]{Time evolution of a slightly perturbed $\tcv{#1}{#2}$ vortex for $\interparam = \num{#3}$, with the values of other parameters taken from set~$\parameterset{2}$. Panels~{#8} show the number density {#6} of atoms in component~1 at the indicated time instants, while panels~{#9} are for the number density {#7} of atoms in component~2. The colorbar scale is in units of $\osclength^{-2}$, where $\osclength = \SI[mode=math]{1.94}{\micro\meter}$. The corresponding stationary state has $\max_{\qpwinding,\qpindexa}\mathrm{Im}(\omega_\qpindexa^{(\qpwinding\mkern+1mu)}/\omega_1)=\num{#5}$ and $\abs{\maxqpwinding}=\num{#4}$. For a video of this evolution, see Supplemental Material~\cite{supplemental_videos}.}

\newcommand{\imaginarypartscaption}[2]{Positive imaginary parts of the excitation frequencies as a function of the intercomponent scattering length $a_{12}$ for the $\tcv{#1}{#2}$~vortex in set~$\parameterset{2}$.}

\begin{document}
\title{Splitting of singly and doubly quantized composite vortices in two-component Bose--Einstein condensates}
\date{\today}
\author{Pekko Kuopanportti}\email{pekko.kuopanportti@gmail.com}
\affiliation{Department of Physics, University of Helsinki, P.O. Box 43, 00014 Helsinki, Finland}
\affiliation{School of Physics and Astronomy, Monash University, Victoria 3800, Australia}
\author{Soumik Bandyopadhyay}
\affiliation{Physical Research Laboratory, Navarangpura, Ahmedabad 380009, Gujarat, India}
\affiliation{Indian Institute of Technology Gandhinagar, Palaj, Gandhinagar 382355, Gujarat, India}
\author{Arko Roy}
\affiliation{INO-CNR BEC Center and Dipartimento di Fisica, Universit\`{a} di Trento, 38123 Trento, Italy}
\affiliation{Max-Planck-Institut f\"{u}r Physik komplexer Systeme, N\"{o}thnitzer Stra{\ss}e 38, 01187 Dresden, Germany}
\author{D. Angom}
\affiliation{Physical Research Laboratory, Navarangpura, Ahmedabad 380009, Gujarat, India}
\begin{abstract}
We study numerically the dynamical instabilities and splitting of singly and doubly quantized composite vortices in two-component Bose--Einstein condensates harmonically confined to quasi two dimensions. In this system, the vortices become pointlike composite defects that can be classified in terms of an integer pair $\tcv{\windingnumber_1}{\windingnumber_2}$ of phase winding numbers. Our simulations based on zero-temperature mean-field theory reveal several vortex splitting behaviors that stem from the multicomponent nature of the system and do not have direct counterparts in single-component condensates. By calculating the Bogoliubov excitations of stationary axisymmetric composite vortices, we find nonreal excitation frequencies (i.e., dynamical instabilities) for the singly quantized $\tcv{1}{1}$ and $\tcv{1}{-1}$ vortices and for all variants of doubly quantized vortices, which we define by the condition $\max_{j\mkern+2mu\in\mkern+2mu\{1,\mkern+2mu 2\}}\mkern+1mu\abs{\windingnumber_j}=2$. While the short-time predictions of the linear Bogoliubov analysis are confirmed by direct time integration of the Gross--Pitaevskii equations of motion, the time integration also reveals intricate long-time decay behavior not captured by the linearized dynamics. First, the $\tcv{1}{\pm 1}$ vortex is found to be unstable against splitting into a $\tcv{1}{0}$ vortex and a $\tcv{0}{\pm 1}$ vortex. Second, the $\tcv{2}{1}$ vortex exhibits a two-step decay process in which its initial splitting into a $\tcv{2}{0}$ vortex and a $\tcv{0}{1}$ vortex is followed by the off-axis splitting of the $\tcv{2}{0}$ vortex into two $\tcv{1}{0}$ vortices. Third, the $\tcv{2}{-2}$ vortex is observed to split into a $\tcv{-1}{1}$ vortex, three $\tcv{1}{0}$ vortices, and three $\tcv{0}{-1}$ vortices. Each of these splitting processes is the dominant decay mechanism of the respective stationary composite vortex for a wide range of intercomponent interaction strengths and relative populations of the two condensate components and should be amenable to experimental detection. Our results contribute to a better understanding of vortex physics, hydrodynamic instabilities, and two-dimensional quantum turbulence in multicomponent superfluids.
\end{abstract}
\preprint{DOI: \href{https://doi.org/10.1103/PhysRevA.100.033615}{10.1103/PhysRevA.100.033615}}
\maketitle

\section{\label{sec:intro}Introduction}

Quantized vortices are $(D-2)$-dimensional topological defects in a spatially $D$-dimensional system that exhibits long-range quantum phase coherence; here $D\in\left\{2,3\right\}$. These quantum whirlpools have been studied in a variety of different systems and branches of physics, such as helium superfluids~\cite{Don1991.book.Vortices}, superconductors~\cite{Par1969.book.superconductivity}, neutron stars~\cite{And1975.Nat256.25}, cosmology~\cite{Vil1994.book.cosmic_strings}, and optics~\cite{Swa1992.PRL69.2503}. Their creation and observation in Bose--Einstein condensates~(BECs) of dilute atomic gases in 1999~\cite{Mat1999.PRL83.2498} and the subsequent detection of regular vortex lattices~\cite{Mad2000.JMO47.2715,Abo2001.Sci292.476,Ram2001.PRL87.210402} were important demonstrations of the superfluidity of the gaseous condensates. Since then, the study of vortices in dilute BECs has flourished~\cite{Fet2009.RMP81.647,And2010.JLTP161.574}, not least because the highly controllable, state-of-the-art BEC experiments now allow for the vortices to be directly imaged and their motion tracked with good spatial and temporal resolution~\cite{Nee2010.PRL104.160401,Fre2010.Sci329.1182,*[{for a comparison of the experimental data with simulations, see }]Kuo2011.PRA83.011603,Mid2011.PRA84.011605,Nav2013.PRL110.225301,Wil2015.PRA91.023621}. Owing to these unique possibilities, vortices in BECs have recently been investigated very actively in the context of two-dimensional quantum turbulence~\cite{Nee2013.PRL111.235301,Whi2014.PNAS111.4719,Kwo2014.PRA90.063627,Seo2017.SciRep7.4587,Gau2019.Sci364.1264,Joh2019.Sci364.1267}. 

In principle, a quantized vortex in a BEC can carry any integer number of circulation quanta. It is well known, however, that a vortex for which this winding number is greater than unity typically has a higher energy than a cluster of single-quantum vortices with the same total circulation. Consequently, such multiply quantized vortices have a tendency to split into single-quantum vortices, which is a manifestly nonlinear phenomenon that has been the subject of interest in both theoretical~\cite{Mot2003.PRA68.023611,Kaw2004.PRA70.043610,Huh2006.PRL97.110406,Mat2006.PRL97.180409,Gaw2006.JPhysB39.L225,Lun2006.PRA74.063620,Kar2009.JPhysB42.095301,Kuo2010.PRA81.023603,Kuo2010.PRA81.033627,Kuo2010.JLTP161.561,Li2012.PRA86.023628,Rab2018.PRA98.023624} and experimental~\cite{Shi2004.PRL93.160406,Iso2007.PRL99.200403,Kuw2010.JPSJ79.034004,Shi2011.JPhysB44.075302} investigations. Moreover, several works have addressed using the splitting of a multiquantum vortex with a sufficiently large winding number (i.e., a ``giant'' vortex) as a means to generate quantum turbulence with nonzero total circulation~\cite{Abr1995.PRB52.7018,Ara1996.PRB53.75,San2014.arXiv.1405.0992,Tel2015.arXiv.1505.00616,Cid2016.PRA93.033651,Cid2017.PRA96.023617}. Besides being interesting due to their splitting tendency, multiply quantized vortices could be used as they are to realize bosonic quantum Hall states~\cite{Ron2011.SciRep1.43} or implement a ballistic quantum switch~\cite{Mel2002.Nat415.60}. 

The aforementioned studies of vortex splitting~\cite{Mot2003.PRA68.023611,Kaw2004.PRA70.043610,Huh2006.PRL97.110406,Mat2006.PRL97.180409,Gaw2006.JPhysB39.L225,Lun2006.PRA74.063620,Kar2009.JPhysB42.095301,Kuo2010.PRA81.023603,Kuo2010.PRA81.033627,Kuo2010.JLTP161.561,Li2012.PRA86.023628,Rab2018.PRA98.023624,Shi2004.PRL93.160406,Iso2007.PRL99.200403,Kuw2010.JPSJ79.034004,Shi2011.JPhysB44.075302} were conducted for a solitary scalar BEC, which is described by a single $\mathbb{C}$-valued order parameter. Perhaps not surprisingly, vortex physics becomes much more diverse when multiple, say $\componentnumber \in\left\{2,3,\myDots\right\}$, scalar condensates come into contact, interact with one another, and thereby constitute a $\componentnumber$-component BEC described by a $\mathbb{C}^\componentnumber$-valued order parameter. Already the simplest multicomponent system, the two-component BEC corresponding to $\componentnumber=2$, supports several stable vortex structures not encountered in single-component BECs, such as coreless vortices~\cite{Mat1999.PRL83.2498}, square vortex lattices~\cite{Mue2002.PRL88.180403,Sch2004.PRL93.210403,Kuo2012.PRA85.043613}, serpentine vortex sheets~\cite{Kas2009.PRA79.023606}, triangular lattices of vortex pairs~\cite{Kuo2012.PRA85.043613}, skyrmions~\cite{Yan2008.PRA77.033621,Mas2011.PRA84.033611,Kuo2015.PRA91.043605}, and meron pairs~\cite{Kas2004.PRL93.250406,Kas2005.PRA71.043611}. Although presently only the coreless vortices~\cite{Mat1999.PRL83.2498} and square vortex lattices~\cite{Sch2004.PRL93.210403} from this list have been verified experimentally, investigation of exotic vortex configurations in two-component BECs is becoming more and more within reach of state-of-the-art experiments. To date, production of two-component BECs has been demonstrated in systems involving either two distinct elements~\cite{Fer2002.PRL89.053202,Mod2002.PRL89.190404,Tha2008.PRL100.210402,Aik2009.NJP11.055035,Cat2009.PRL103.140401,Car2011.PRA84.011603,Ler2011.EPJD65.3,Pas2013.PRA88.023601,Wac2015.PRA92.053602,Wan2016.JPhysB49.015302,Bur2018.PRA98.063616,Cas2019.LaserPhysLett16.035501}, two different isotopes of the same element~\cite{Pap2008.PRL101.040402,Sug2011.PRA84.011610,Ste2013.PRA87.013611}, or two different spin states of the same isotope~\cite{Mya1997.PRL78.586,Hal1998.PRL81.1539,Mat1999.PRL83.2498,Del2001.PRA63.051602,Sch2004.PRL93.210403,And2009.PRA80.023603,Cab2018.Sci359.301,Sem2018.PRL120.235301}.

Besides the vast array of static vortex structures listed above, two-component BECs also exhibit more intricate vortex dynamics than the single-component system. This is certainly the case for the splitting of vortices, already because each vortex must then be characterized by two integer winding numbers instead of just one, leading to composite vortices that have no single-component counterparts. Apart from a few earlier studies, the stability properties and splitting dynamics of such vortices remain largely unknown. The excitation spectra and the related instabilities of axisymmetric vortex states in harmonically trapped two-component BECs were examined by Skryabin~\cite{Skr2000.PRA63.013602}, but only for cases where just one of the components contains a vortex. Ishino, Tsubota, and Takeuchi~\cite{Ish2013.PRA88.063617} studied the stability and splitting of so-called counter-rotating vortex states, in which the two components host vortices of equal but opposite winding numbers, and found exotic splitting patterns where each of the two vortices splits into both vortices and antivortices. Ishino \emph{et al.}, however, did not consider other types of composite vortices and limited their investigation to number-balanced systems and a few selected interaction strengths. The stability of singly quantized composite vortices, for which each component carries at most one circulation quantum, was studied in Refs.~\cite{Gar2000.PRL84.4264,Per2000.PRA62.033601,Yak2009.PRA79.043629,Brt2010.PRA82.053610}. In addition, the related but distinct problem of dynamical instabilities of coreless vortices in spin-1 BECs (for which $\componentnumber = 3$) was considered in Refs.~\cite{Pie2007.PRA76.023610,Tak2009.PRA79.023618}. 

Motivated by the limited amount of existing literature relative to the expected richness of the phenomenon, we investigate here the splitting dynamics and underlying dynamical instabilities of axisymmetric singly and doubly quantized composite vortices in harmonically trapped two-component BECs using a broader set of system parameters than in Refs.~\cite{Skr2000.PRA63.013602,Ish2013.PRA88.063617}. Our simulations based on the Gross--Pitaevskii~(GP) and Bogoliubov equations reveal vortex splitting behaviors that do not appear for multiply quantized vortices in single-component BECs. First, we detect dynamical splitting instabilities even for vortex states where neither component hosts a multiquantum vortex. Second, some dynamically unstable composite vortices are observed to exhibit a split-and-revival phenomenon that involves splitting and subsequent recombination of their constituent vortices. Third, we find that a doubly quantized vortex in one component, when accompanied by an oppositely charged vortex in the other component, tends to split into three single-quantum vortices and one single-quantum antivortex. This threefold-symmetric splitting pattern is in drastic contrast to a doubly quantized vortex in a single-component BEC, which can only split into two single-quantum vortices of the same sign. Our results demonstrate that this peculiar splitting mode is dominant in an extensive region of the system parameter space, can be visually identified by counting the emerging vortices, and should therefore be amenable to experimental verification.

The remainder of this paper is organized as follows. In Sec.~\ref{sec:theory}, we review the zero-temperature mean-field theory of the two-component BECs and describe how it is employed in our simulations. Section~\ref{sec:results} presents our numerical results on the dynamical instabilities and splitting of singly and doubly quantized composite vortices. In Sec.~\ref{sec:discussion}, we summarize our main findings, discuss their implications and limitations, and suggest possible future extensions.

\section{\label{sec:theory}Theory and methods}
\subsection{\label{subsec:mft}Gross--Pitaevskii model and composite vortices}

Our theoretical treatment starts from the coupled time-dependent GP equations for two BECs in an axisymmetric harmonic trap~\cite{Kas2005.IJMPB19.1835}:
\begin{equation}\label{eq:2dgpe}
\begin{split}
 i\hbar \frac{\partial}{\partial t} \Psi_j(r,\phi,t)=
\bigg\{&-\frac{\hbar^2}{2m_j}\left[\frac{1}{r}\frac{\partial}{\partial r}\left(r \frac{\partial}{\partial r}\right)+\frac{1}{r^2}\frac{\partial^2}{\partial \phi^2}\right]+\frac{1}{2}m_j\omega_j^2 r^2\\&+ \intrag{j} \abs{\Psi_j}^2+g_{12}\abs{\Psi_{3-j}}^2 \bigg\} \Psi_j(r,\phi,t),
\end{split}
\end{equation}
where $\pointcoord{r}{\phi}$ are the polar coordinates, $j\in\left\{1,2\right\}$, and the order-parameter fields $\Psi_j$ are normalized such that $\norm{\Psi_j} \coloneqq \left[ \iint \abs{\Psi_j\pointcoord{r}{\phi}}^2 \intmeasure\right]^{1/2} =N_j^{1/2}$. Here $N_j$, $m_j$, and $\omega_j$ denote, respectively, the total number, the mass, and the radial harmonic trapping frequency of atoms of component $j$. For simplicity, we limit our attention to quasi-two-dimensional configurations, which correspond to, e.g., highly oblate traps with strong axial confinement that renders $\Psi_1$ and $\Psi_2$ approximately Gaussian in the axial direction. The intracomponent interaction strengths $\intrag{1}$ and $\intrag{2}$ are assumed to be positive, whereas for the intercomponent parameter $g_{12}$ we consider all values such that $g_{12}^2 < \intrag{1}\intrag{2}$, which defines the so-called miscible regime.

In this paper, we are interested in the instabilities and splitting of axisymmetric vortex states. To this end, we seek stationary solutions to Eqs.~\eqref{eq:2dgpe} of the form
\begin{equation}\label{eq:axisymm_ansatz}
\Psi_j\left(r,\phi,t\right) = \amplitude_j\left(r\right) e^{i\windingnumber_j\phi} e^{-i\mu_j t/\hbar},
\end{equation}
where $\windingnumber_j\in\mathbb{Z}$ denotes the phase winding number, i.e., the charge, of the central vortex in component $j$; the chemical potentials $\mu_j$ are used to enforce the normalization conditions $\norm{\amplitude_j}^2 = N_j$. With this ansatz, Eqs.~\eqref{eq:2dgpe} reduce to two coupled nonlinear ordinary differential equations for the radial functions $\amplitude_1$ and $\amplitude_2$:
\begin{equation}\label{eq:axisymGPE}
\begin{split}
-\frac{\hbar^2}{2m_j} \left(\frac{\ud^2}{\ud r^2} + \frac{1}{r} \frac{\ud}{\ud r} - \frac{ \windingnumber_j^2}{r^2}\right)\amplitude_j +\frac{1}{2}m_j \omega_j^2 r^2 \amplitude_j & \\+ \intrag{j} \abs{\amplitude_j}^2 \amplitude_j +g_{12} \abs{\amplitude_{3-j}}^2 \amplitude_j &= \mu_j \amplitude_j.
\end{split}
\end{equation}
We shall refer to the lowest-energy solution of Eqs.~\eqref{eq:axisymGPE} for a given winding-number pair (and given system parameters) as a stationary $\tcv{\windingnumber_1}{\windingnumber_2}$ vortex. Here the notion ``$\tcv{\windingnumber_1}{\windingnumber_2}$ vortex'' is defined more generally as a sufficiently pointlike phase defect about which $\mathrm{arg}\left(\Psi_1\right)$ winds by $\windingnumber_1 \times 2 \pi$ and $\mathrm{arg}\left(\Psi_2\right)$ winds by $\windingnumber_2 \times 2 \pi$, with $\abs{\windingnumber_1}+\abs{\windingnumber_2} > 0$~\footnote{Here ``sufficiently pointlike'' means that all the phase singularities involved in the definition should lie within a circle of radius smaller than $\min_j \xi_j$, where $\xi_j$ is the healing length of component $j$. In Eqs.~\eqref{eq:axisymm_ansatz} and~\eqref{eq:axisymGPE}, the stationary $\tcv{\windingnumber_1}{\windingnumber_2}$ vortex is strictly a point defect located exactly at the origin.}. For $\windingnumber\in\mathbb{Z}_+$, we further define a ``$\windingnumber$-quantum composite vortex'' as a $\tcv{\windingnumber_1}{\windingnumber_2}$ vortex for which $\max_j \abs{\windingnumber_j}=\windingnumber$. By a ``coreless vortex,'' in turn, we mean a $\tcv{\windingnumber_1}{0}$ or $\tcv{0}{\windingnumber_2}$ vortex for which the total particle density $\ntot=\sum_j n_j =\sum_j\,\abs{\Psi_j}^2$ does not vanish at the phase singularity. Conversely, a vortex for which $\ntot=0$ at the singularity is classified as ``cored.'' 

\subsection{\label{subsec:bogoliubov}Linear stability analysis}

To study the stability properties of a given stationary $\tcv{\windingnumber_1}{\windingnumber_2}$ vortex, we decompose the order-parameter components as
\begin{equation}\label{eq:decomposition}
\Psi_j(r,\phi,t)=e^{-i\mu_j t/\hbar} e^{i\windingnumber_j\phi} \left[\amplitude_j(r)+\perturbation_{j}(r,\phi,t)\right],
\end{equation}
where the functions $\perturbation_{j}$ are assumed to be small in the sense that $\norm{\perturbation_j}^2 \ll N_j$. By substituting Eqs.~\eqref{eq:decomposition} into Eqs.~\eqref{eq:2dgpe}, omitting the second- and third-order terms in $\perturbation_j$, and seeking solutions in the form
\begin{equation}\label{eq:oscillation}
\perturbation_{j}(r,\phi,t)=\sum_\qpindexa \sum_{\qpwinding\in\mathbb{Z}} \left[ u_{\qpindexa,j}^{(\qpwinding\mkern+1mu)}(r) e^{ i \qpwinding \phi-i\omega_\qpindexa^{(\qpwinding\mkern+1mu)} t}+ v_{\qpindexa,j}^{(\qpwinding\mkern+1mu)\,\ast}(r) e^{ -i \qpwinding \phi+i\omega_\qpindexa^{(\qpwinding\mkern+1mu)\,\ast} t} \right],
\end{equation}
we obtain the Bogoliubov equations
\begin{equation}\label{eq:axisym2cBogoeq}
\bogoper^{(\qpwinding\mkern+1mu)}\bm{w}_\qpindexa^{(\qpwinding\mkern+1mu)}(r) = \hbar\omega_\qpindexa^{(\qpwinding\mkern+1mu)} \bm{w}_\qpindexa^{(\qpwinding\mkern+1mu)}(r),
\end{equation}
where $\bm{w}_\qpindexa^{(\qpwinding\mkern+1mu)} = \left(u_{\qpindexa,1}^{(\qpwinding\mkern+1mu)},u_{\qpindexa,2}^{(\qpwinding\mkern+1mu)},v_{\qpindexa,1}^{(\qpwinding\mkern+1mu)},v_{\qpindexa,2}^{(\qpwinding\mkern+1mu)}\right)^\mathrm{T}$ and
\begin{equation}\label{eq:bogoper}
\bogoper^{(\qpwinding\mkern+1mu)}=\begin{pmatrix} 
\axisymoperator_{1}^{(\qpwinding+\windingnumber_1)} & g_{12}\amplitude_1 \amplitude_2^\ast & \intrag{1}\amplitude_1^2 & g_{12} \amplitude_1 \amplitude_2 \\
g_{12}\amplitude_1^\ast \amplitude_2 & \axisymoperator_{2}^{(\qpwinding+\windingnumber_2)} & g_{12} \amplitude_1 \amplitude_2 & \intrag{2} \amplitude_2^2\\
-\intrag{1}(\amplitude_1^\ast)^2 & -g_{12}\amplitude_1^\ast \amplitude_2^\ast & -\axisymoperator_{1}^{(\qpwinding-\windingnumber_1)} & -g_{12}\amplitude_1^\ast \amplitude_2 \\
-g_{12} \amplitude_1^\ast \amplitude_2^\ast & -\intrag{2}(\amplitude_2^\ast)^2 & -g_{12}\amplitude_1 \amplitude_2^\ast & -\axisymoperator_{2}^{(\qpwinding-\windingnumber_2)}
\end{pmatrix}.
\end{equation}
In Eqs.~\eqref{eq:oscillation}, the integer $\qpwinding$ specifies the angular momentum of the excitation (in units of $\hbar$) with respect to the condensate, and $\qpindexa\in\mathbb{Z}_+$ is an index for the different eigensolutions with the same $\qpwinding$. The diagonal of the matrix operator $\bogoper^{(\qpwinding\mkern+1mu)}$ in Eq.~\eqref{eq:bogoper} consists of the linear differential operators
\begin{equation}\label{eq:diagonaloper}
\begin{split}
\axisymoperator_{j}^{(k)} = &-\frac{\hbar^2}{2m_j} \left[\frac{1}{r}\frac{\ud}{\ud r}\left(r \frac{\ud}{\ud r}\right)-\frac{k^2}{r^2}\right] +\frac{1}{2} m_j \omega_j^2 r^2 -\mu_j\\ &+ 2 \intrag{j}\abs{\amplitude_j}^2 + g_{12} \abs{\amplitude_{3-j}}^2,
\end{split}
\end{equation}
where $k=\qpwinding \pm \windingnumber_j$.

Equations~\eqref{eq:axisym2cBogoeq} can be used to determine the local stability characteristics of the stationary vortex state in question. If there exists an integer $\qpwinding \in\mathbb{Z}$ for which the excitation spectrum $\bigl\{\omega_\qpindexa^{(\qpwinding\mkern+1mu)} \mid \qpindexa\in\mathbb{Z}_+\bigr\}$ contains at least one eigenfrequency $\omega_\qpindexa^{(\qpwinding\mkern+1mu)}$ with a positive imaginary part $\mathrm{Im}(\omega_\qpindexa^{(\qpwinding\mkern+1mu)})>0$, the state is \emph{dynamically unstable}; otherwise, the state is dynamically stable. On the other hand, if for some $\qpwinding$ the spectrum contains an eigenfrequency with a negative real part $\mathrm{Re}(\omega_\qpindexa^{(\qpwinding\mkern+1mu)})<0$ and a non-negative eigenfield pseudonorm $\sum_j \norm{u_{\qpindexa,j}^{(\qpwinding\mkern+1mu)}}^2- \sum_j \norm{v_{\qpindexa,j}^{(\qpwinding\mkern+1mu)}}^2 \geq 0$, the state is \emph{energetically unstable}; otherwise, it is (locally) energetically stable.

As can be observed from Eqs.~(\ref{eq:oscillation}), the amplitudes of excitation modes with $\mathrm{Im}(\omega_\qpindexa^{(\qpwinding\mkern+1mu)})>0$ are predicted to grow exponentially in time; accordingly, small perturbations of a dynamically unstable stationary state tend to result in large changes in its structure. Furthermore, since the first- and second-order contributions of these dynamically unstable excitation modes to the system energy can be shown to vanish~\cite{Mor2000.JPhysB33.3847}, they can become populated and cause the state to decay even in the absence of dissipation. In contrast, populating an energetically unstable excitation mode would reduce the system energy and, therefore, would typically occur only if there were a dissipation mechanism available, such as a non-negligible thermal-gas component. Dissipation could be added phenomenologically to Eqs.~\eqref{eq:2dgpe} by replacing $t$ with $(1-i\Lambda)t$, where $0 < \Lambda \ll 1$ is a dimensionless damping parameter inversely proportional to the energy relaxation time, and by explicitly enforcing the normalization conditions $\norm{\Psi_j}^2 = N_j$ during the time evolution~\cite{Cho1998.PRA57.4057}. In this paper, however, we set $\Lambda=0$ and thus consider only dynamics that conserve the energy, the norms $\norm{\Psi_j}$, and the total axial angular momentum ${L}_z = -i\hbar \sum_j \iint \Psi_j^\ast\,\partial_\phi \Psi_j\, \intmeasure$.

For dynamically unstable multiquantum vortices, in particular, the nonreal excitation frequencies usually indicate instability against splitting of the multiply quantized vortex into singly quantized ones. In the case of a dynamically unstable $\tcv{\windingnumber_1}{\windingnumber_2}$ vortex, the quantity $\max_{\qpwinding}\max_\qpindexa \mathrm{Im}(\omega_\qpindexa^{(\qpwinding\mkern+1mu)}/2\pi)$ and the maximizing angular-momentum quantum number $\qpwinding$ can be used to estimate, respectively, the inverse lifetime of the vortex and the order of rotational symmetry of its typical splitting pattern~\cite{Mot2003.PRA68.023611,Kuo2010.PRA81.033627,Rab2018.PRA98.023624}. It should be noted, however, that the exponentially growing modes rapidly drive the system away from the linear regime of Eqs.~\eqref{eq:decomposition}--\eqref{eq:diagonaloper}; consequently, the long-time dynamics of dynamically unstable states must instead be described with the time-dependent GP equations~\eqref{eq:2dgpe}.

\subsection{\label{subsec:time_evolution}Time-evolution simulations}

The Bogoliubov stability analysis cannot be used to draw rigorous conclusions about the behavior of perturbed stationary states far from the limit of infinitesimal perturbations; this is a particularly serious restriction for dynamically unstable states. Therefore, to go beyond the linear-response regime, we simulate the full two-dimensional dynamics of the unstable composite vortices by directly integrating the time-dependent GP equations~\eqref{eq:2dgpe}. The time integration is performed with a split-step Crank--Nicolson method~\cite{Mur2009.ComputPhysCommun180.1888,Vud2012.ComputPhysCommun183.2021} adapted for two-component BECs~\cite{Roy2018.arXiv.1806.01244}. To obtain a convenient initial state, we first propagate Eqs.~\eqref{eq:2dgpe} in imaginary time, which formally corresponds to replacing $t$ with $-i\tau$, where $\tau > 0$. At each imaginary-time step, we apply the transformation
\begin{equation}
\Psi_{j}(r, \phi) \replacedwith \frac{N_j^{1/2}}{\norm{\Psi_{j}}} \abs{\Psi_{j}(r, \phi)}{\rm e}^{i \windingnumber_{j}\phi},
\end{equation}
which ensures proper normalization and introduces a vortex of charge $\windingnumber_{j}$ into the center of component $j$. The imaginary-time evolution is continued until the solution has approximately converged. The resulting near-equilibrium state, which can be viewed as comprising the stationary $\tcv{\windingnumber_1}{\windingnumber_2}$ vortex and a small-amplitude perturbation as in Eqs.~\eqref{eq:decomposition}, is then used as the initial ($t=0$) state for Eqs.~\eqref{eq:2dgpe} and propagated forward in real time (i.e., without the phenomenological damping term) until the composite vortex has decayed. The time-evolution calculations are performed using a square spatial grid of $\num{251}\times\num{251}$ points with a grid spacing of $\Delta x = \Delta y = \num{0.05}\,\osclength$ and a time step of $\Delta t = \num{d-4}/\omega_1$.

\subsection{\label{subsec:parametrization}Parametrization}

In the numerics, we cast Eqs.~\eqref{eq:2dgpe},~\eqref{eq:axisymGPE}, and~\eqref{eq:axisym2cBogoeq} into dimensionless form by measuring energy, time, and length in units of $\hbar\omega_1$, $\omega_1^{-1}$, and $\osclength=\sqrt{\hbar/m_1\omega_1}$, respectively, and normalizing the dimensionless order parameters to unity. This results in a model that is fully specified by the vortex winding numbers $\tcv{\windingnumber_1}{\windingnumber_2}\in\mathbb{Z}^2$ and six dimensionless real parameters, namely, $\omega_2/\omega_1$, $m_2/m_1$, $\interparam\coloneqq g_{12}/\sqrt{\intrag{1}\intrag{2}}$, $\polarization\coloneqq(N_1-N_2)/(N_1+N_2)$, $\intrag{2}/\intrag{1}$, and $\geff\coloneqq m_1(N_1 \intrag{1} + N_2 g_{12})/\hbar^2$~\footnote{In addition to the winding-number pair $\tcv{\windingnumber_1}{\windingnumber_2}$, actually only five independent real parameters are needed to specify Eqs.~\eqref{eq:axisymGPE} fully, and thus parametrize the stationary $\tcv{\windingnumber_1}{\windingnumber_2}$ vortices themselves, but all six are needed in Eqs.~\eqref{eq:axisym2cBogoeq} to determine the excitation spectra of said vortices.}. To enable a systematic numerical investigation of the model, we will reduce the number of free parameters as follows.

Throughout the paper, we set $\omega_1=\omega_2$ and $m_1=m_2$, corresponding physically to a two-component BEC formed from two different spin states of the same atomic isotope. For the winding numbers $\windingnumber_1$ and $\windingnumber_2$, we assume $\windingnumber_1 \in \left\{1,2\right\}$ and $\abs{\windingnumber_2}\leq \windingnumber_1$, which amounts to limiting the investigation to all truly different singly and doubly quantized composite vortices, of which there are three and five variants, respectively~\footnote{For each $\windingnumber\in\mathbb{Z}_+$, there are $2\windingnumber+1$ truly different combinations of winding numbers that satisfy the definition of a $\windingnumber$-quantum composite vortex given at the end of Sec.~\ref{subsec:mft}.}.

For the remaining four parameters ($\interparam$, $\polarization$, $\intrag{2}/\intrag{1}$, and $\geff$), we use two different sets. The first parameter set, which we refer to as $\parameterset{1}$, is used for the linear stability analysis to obtain vortex stability diagrams in the parameter plane of the relative intercomponent interaction strength $\interparam$ and the ``polarization'' $\polarization$ for each winding-number pair under consideration. Here both $\interparam$ and $\polarization$ are varied in the open interval $\openinterval{-1}{1}$, which for $\interparam$ corresponds to the miscible regime. These two parameters are taken as the ``phase-space variables'' because they are the most relevant to our purpose of studying how the coupling between the two components affects the composite-vortex stability. The parameter $\intrag{2}/\intrag{1}$, on the other hand, is scaled in $\parameterset{1}$ such that both components have the same radius in the Thomas--Fermi approximation in the absence of vortices~\cite{Ho1996.PRL77.3276,Pol2015.PRA91.053626,Kuo2019.JPhysB52.015001}, yielding the relation
\begin{equation}\label{eq:g2perg1}
\frac{\intrag{2}}{\intrag{1}}=\frac{\left[\sqrt{1-\left(1-\interparam^2\right)\polarization^2}-\polarization\interparam \right]^2}{\left(1-\polarization\right)^2}.
\end{equation}
When Eq.~\eqref{eq:g2perg1} holds, the vortex-free Thomas--Fermi radius is determined by the effective overall interaction strength $\geff$, which in $\parameterset{1}$ is fixed at \num{1000}. As a result, in $\parameterset{1}$ both BEC components have an outer radius approximately equal to $R=\osclength \sqrt[4]{4\geff/\pi}=\num{5.97}\,\osclength$, independent of the values of $\interparam$ and $\polarization$. We have confirmed that the axisymmetric vortex-free state $\tcv{\windingnumber_1}{\windingnumber_2}=\tcv{0}{0}$ of $\parameterset{1}$ 
is energetically and dynamically stable in the whole domain $\pointcoord{\interparam}{\polarization}\in{\openinterval{-1}{1}}^2$, which indicates that any instabilities appearing in our stability diagrams are due to the presence of the vortices~\footnote{Note, however, that the lowest eigenfrequency associated with a positive pseudonorm approaches zero from above as $\interparam\nearrow 1$, signaling the emergence of the segregation instability in the immiscible regime $\interparam > 1$. At the end of Sec.~\ref{sec:discussion}, we also comment on the opposite limit $\interparam \searrow -1$}.

The second parameter set, $\parameterset{2}$, is used in all the time-evolution simulations and is designed to clarify the connection to BEC experiments. To this end, we choose $\parameterset{2}$ according to the already realized two-component BEC of the $^{87}$Rb hyperfine states $\ket{F = 1, m_F = -1} \eqqcolon \ket{1}$ and $\ket{F = 2, m_F = +1} \eqqcolon \ket{2}$~\cite{Hal1998.PRL81.1539,Mer2007.PRL99.190402}. We assume a harmonic potential with radial trap frequencies $\omega_1 = \omega_2 = 2\pi\times\SI[mode=math]{30.83}{\hertz}$~\cite{Mer2007.PRL99.190402} and set the ratios of the axial to radial trapping frequencies to $\omega_{z,1}/\omega_1 = \omega_{z,2}/\omega_2 = \num{40.0}$, which ensures that $\mu_{j} \ll \hbar\omega_{z,j}$ and renders the system quasi-two-dimensional. The harmonic oscillator length for this configuration is $\osclength =\SI[mode=math]{1.94}{\micro\meter}$. For the total numbers of atoms, we take the representative values $N_1 = N_2 = \num{2d3}$, implying that $P\coloneqq (N_1 - N_2)/(N_1 + N_2) = 0$. The intracomponent scattering lengths $\intraa{1}$ and $\intraa{2}$ are fixed at $\num{100.4}\,\bohrradius$ and $\num{95.44}\,\bohrradius$~\cite{Ego2013.PRA87.053614}, respectively, where $\bohrradius$ is the Bohr radius. On the other hand, the intercomponent scattering length $a_{12}$ of this mixture can be varied with a magnetic Feshbach resonance~\cite{Mar2002.PRL89.283202,Erh2004.PRA69.032705,Toj2010.PRA82.033609}, which motivates us to use several values for the intercomponent interaction strength $\interparam \coloneqq g_{12}/\sqrt{\intrag{1}\intrag{2}} = a_{12}/\sqrt{\intraa{1}\intraa{2}}$~\footnote{In our parameter set~$\parameterset{2}$, the dimensionless interaction strength $\geff$ is given by the formula $\geff = \num{173.53} + \num{169.19}\times\interparam$.}. Despite its specific nature, set~$\parameterset{2}$ turns out to be general enough to exhibit the main features of composite-vortex splitting that we encounter in $\parameterset{1}$. We expect the conclusions drawn from $\parameterset{1}$ and $\parameterset{2}$ to apply qualitatively to other kinds of two-component BECs as well.

\section{\label{sec:results}Results}

In this section, we present and analyze the stability diagrams of singly~(Sec.~\ref{subsec:single-quantum_results}) and doubly~(Sec.~\ref{subsec:two-quantum_results}) quantized composite vortices and illustrate the corresponding dynamics with representative examples from the time-evolution simulations (see also Supplemental Material~\cite{supplemental_videos}). To obtain a pair of stability diagrams for each winding-number combination $\tcv{\windingnumber_1}{\windingnumber_2}$, we first find the stationary axisymmetric vortex states by solving Eqs.~\eqref{eq:axisymGPE} over the two-dimensional parameter space $\pointcoord{\interparam}{\polarization}\in{\openinterval{-1}{1}}^2$, with the values of the other parameters taken from set~$\parameterset{1}$. We then solve the Bogoliubov equations~\eqref{eq:axisym2cBogoeq} for each stationary solution over all relevant values of $\qpwinding \in\mathbb{Z}$ and determine, in particular, the magnitude of the dominant dynamical instability, $\max_{\qpwinding}\max_\qpindexa \mathrm{Im}(\omega_\qpindexa^{(\qpwinding\mkern+1mu)})$, and the angular-momentum quantum number $\qpwinding=\maxqpwinding$ for which this maximum occurs. Although in principle we should consider all integer values of $\qpwinding$, numerical evidence indicates that it suffices to check only the cases $\abs{\qpwinding\mkern+1mu} \leq 5$. Moreover, due to the symmetries of Eqs.~\eqref{eq:axisym2cBogoeq}, only the absolute value of $\maxqpwinding$ is significant. The various vortex-splitting behaviors encountered in the resulting two-dimensional diagrams are then illustrated with the corresponding time series generated by the GP equations~\eqref{eq:2dgpe} for the experimentally motivated parameter set~$\parameterset{2}$. Although we have systematically performed this analysis for all three variants of single-quantum composite vortices and all five variants of two-quantum composite vortices, brevity compels us to leave some of the less interesting results to Appendices~\ref{app:single-quantum_results} and~\ref{app:two-quantum_results}.

\subsection{\label{subsec:single-quantum_results}Singly quantized composite vortices}

We begin with a brief account of our results for the singly quantized composite vortices $\tcv{\windingnumber_1}{\windingnumber_2}=\tcv{1}{0}$, $\tcv{1}{1}$, and $\tcv{1}{-1}$; we refer to Appendix~\ref{app:single-quantum_results} for the associated numerical data. The stability properties of these vortices may be compared with those of the stationary axisymmetric solitary vortex of winding number $\windingnumber=1$ in a harmonically trapped, quasi-two-dimensional single-component BEC: Such a vortex state is known to be dynamically stable but energetically unstable at all nonzero values of the interaction-strength parameter, with the energetic instability supplied by the so-called anomalous mode, which has $\qpwinding=-1$ and corresponds to the vortex spiraling away from the trap center in a counterclockwise direction. No energetic instabilities exist for other values of $\qpwinding$. 

While we find the $\tcv{1}{0}$ vortex of our two-component system to be dynamically stable in the entire parameter set~$\parameterset{1}$, this is not true for the $\tcv{1}{1}$ and $\tcv{1}{-1}$ vortices: both of them exhibit formidable dynamical instabilities for $\abs{\qpwinding\mkern+1mu}=1$ over extensive regions of the $\interparam\mkern-1.5mu\polarization$ parameter space. These instabilities can be regarded as describing the splitting of the $\tcv{1}{\pm 1}$ composite vortex into a $\tcv{1}{0}$ vortex and a $\tcv{0}{\pm 1}$ vortex. For the $\tcv{1}{1}$ vortex, the dynamical instabilities exist only for repulsive intercomponent interactions, $\interparam>0$, whereas the $\tcv{1}{-1}$ vortex can be dynamically unstable also for $\interparam < 0$. These findings are consistent with Ref.~\cite{Ish2013.PRA88.063617}. It should be noted that although the $\tcv{1}{0}$ vortex is dynamically stable in $\parameterset{1}$, it does support dynamical instabilities with $\abs{\qpwinding\mkern+1mu}=1$ outside this parameter set, as shown in Refs.~\cite{Gar2000.PRL84.4264,Per2000.PRA62.033601}; these instabilities, however, are significantly weaker than those obtained for the $\tcv{1}{1}$ and $\tcv{1}{-1}$ vortices.

For the $\tcv{1}{0}$ and $\tcv{1}{1}$ vortices, energetic instabilities only occur for $\qpwinding=-1$, while for the $\tcv{1}{-1}$ vortex they can exist for both $\qpwinding=-1$ and $1$. Energetically unstable but dynamically stable vortices are observed not to decay under the small initial perturbations and conservative dynamics considered in this paper, but are expected to do so when dissipation is added using, for example, the phenomenological damping term mentioned in Sec.~\ref{subsec:bogoliubov}. Interestingly, the $\tcv{1}{0}$ and $\tcv{1}{-1}$ vortices also show narrow regions of energetic stability in the $\interparam\mkern-1.5mu\polarization$ plane. Such states are predicted to be robust against small perturbations even in the presence of dissipation.

\subsection{\label{subsec:two-quantum_results}Doubly quantized composite vortices}

\begin{figure}[b]
\includegraphics[width=1.0\columnwidth,keepaspectratio]{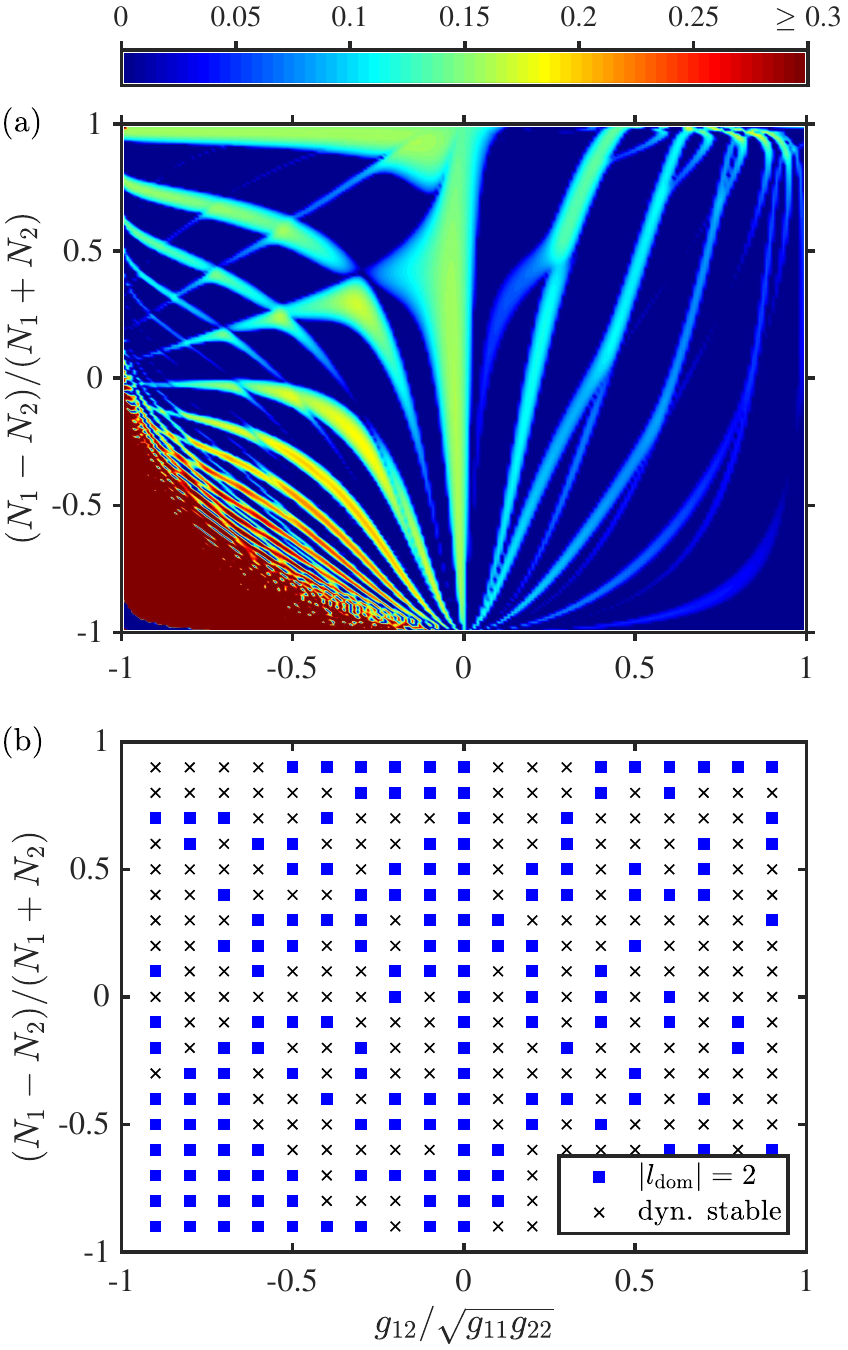}
\caption{(a) Maximum imaginary part of the excitation frequencies $\{\omega_\qpindexa^{(\qpwinding\mkern+1mu)}\}$ of the stationary $\tcv{2}{0}$ vortex as a function of the relative intercomponent interaction strength $\interparam\coloneqq g_{12}/\sqrt{\intrag{1}\intrag{2}}$ and the particle-number polarization $\polarization\coloneqq(N_1-N_2)/(N_1+N_2)$. The colorbar scale is in units of $\omega_1$. (b)~The angular-momentum quantum number $l$ of the excitation that yields the maximum imaginary part at the given point $\pointcoord{\interparam}{\polarization}$. This excitation corresponds to the dominant dynamical instability of the vortex state in question; accordingly, we denote the maximizing $\qpwinding$ value by $\maxqpwinding$. All dynamically stable states in panel~(b), which satisfy $\{\omega_\qpindexa^{(\qpwinding\mkern+1mu)}\}\subseteq \mathbb{R}$, are energetically unstable (i.e., they support a negative-frequency excitation that has a positive pseudonorm). The other parameters in Eqs.~\eqref{eq:axisymGPE} are chosen such that $\omega_1 = \omega_2$, $m_1=m_2$, $\intrag{2}/\intrag{1}=[\sqrt{1-\left(1-\interparam^2\right)\polarization^2}-\polarization\interparam ]^2/\left(1-\polarization\right)^2$, and $\geff \coloneqq m_1\left( N_1 \intrag{1} + N_2 g_{12}\right)/\hbar^2=\num{1000}$, yielding the vortex-free Thomas--Fermi radii of $R_1=R_2= \num{5.97}\times\sqrt{\hbar/m_1\omega_1}$ for all states under consideration. We refer to this parameter scaling as set~$\parameterset{1}$.}\label{fig:maxim-2-0}
\end{figure}

Let us now set $\windingnumber_1=2 \geq \abs{\windingnumber_2}$ and consider the five different types of doubly quantized composite vortices in our two-component system: $\tcv{\windingnumber_1}{\windingnumber_2}=\tcv{2}{0}$, $\tcv{2}{\pm 1}$, and $\tcv{2}{\pm 2}$. In the harmonically trapped, quasi-two-dimensional single-component BEC, the stability of the stationary axisymmetric two-quantum vortex is known to be a quasiperiodic function of the dimensionless interaction strength $m_1 N_1 \intrag{1}/\hbar^2$, the vortex state being either dynamically stable but energetically unstable (due to excitations with $\qpwinding=-2$ or $-1$) or dynamically unstable. The dynamical instability can only occur for $\abs{\qpwinding\mkern+1mu}=2$ and corresponds to a twofold-symmetric, linear-chain splitting instability observed in various experiments on multiply quantized vortices~\cite{Shi2004.PRL93.160406,Iso2007.PRL99.200403,Kuw2010.JPSJ79.034004}.

The case $\tcv{\windingnumber_1}{\windingnumber_2}=\tcv{2}{0}$ is classified as a doubly quantized coreless vortex. Figure~\ref{fig:maxim-2-0} shows its stability diagrams, namely, the magnitude of the dominant dynamical instability, $\max_{\qpwinding}\max_\qpindexa \mathrm{Im}(\omega_\qpindexa^{(\qpwinding\mkern+1mu)})$, and the angular-momentum quantum number $\maxqpwinding$ for which this maximum occurs, in the $\interparam\mkern-1.5mu\polarization$ parameter plane. Here $\interparam\coloneqq g_{12}/\sqrt{\intrag{1}\intrag{2}}$ is the relative intercomponent interaction strength and $\polarization\coloneqq(N_1-N_2)/(N_1+N_2)$ is the particle-number polarization. The $\tcv{2}{0}$ vortex is observed to behave qualitatively similarly to the two-quantum vortex in the single-component BEC~\footnote{In particular, on the vertical line $\interparam=0$ the two components are decoupled, and the dominant dynamical instability reduces exactly to that of a two-quantum vortex in a single-component BEC with a dimensionless coupling constant $m_1 \intrag{1} N_1 /\hbar^2=\num{1000}$, yielding $\abs{\maxqpwinding}=\num{2}$ and $\max_{\qpwinding,\qpindexa} \mathrm{Im}(\omega_\qpindexa^{(\qpwinding\mkern+1mu)}/\omega_1)=\num{0.137}$ for all $\pointcoord{\interparam}{\polarization}=\pointcoord{0}{\polarization}$, $ -1 < \polarization < 1$.}, either being dynamically stable (but energetically unstable) or exhibiting a dynamical instability for $\abs{\qpwinding\mkern+1mu}=2$. As is evident from Fig.~\ref{fig:maxim-2-0}(a), the repulsive intercomponent interaction tends to stabilize the $\tcv{2}{0}$ vortex, with the vortex-free second component acting as an effective pinning potential for the vortex in the first component when $\interparam >0$. The stabilizing effect becomes more pronounced with decreasing $\polarization$; this is consistent with the linear dependence of the average height of the effective potential $\intrag{2}\abs{\Psi_2}^2$ on $N_2$. Note that the tendency to stabilize with increasing $\interparam$ is to be expected for all coreless vortices $\tcv{\windingnumber}{0}$ and $\tcv{0}{\windingnumber}$, where $0\neq \windingnumber\in\mathbb{Z}$.

\begin{figure}[t]
\includegraphics[width=1.0\columnwidth,keepaspectratio]{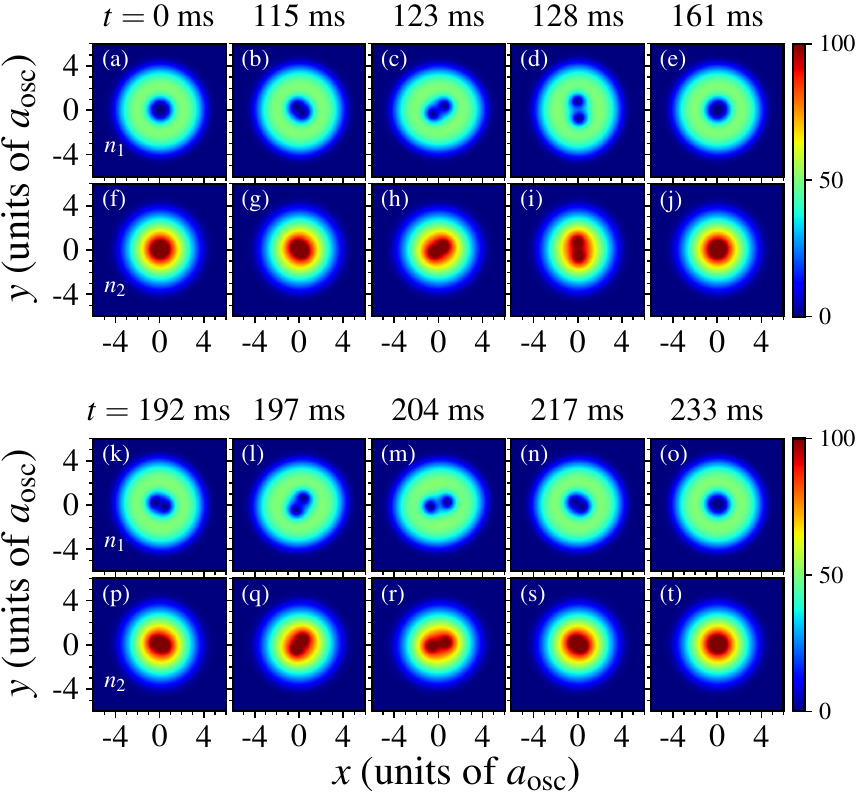}
\caption{\timeseriescaption{2}{0}{0.7}{2}{0.0903}{$n_1\coloneqq \abs{\Psi_1}^2$}{$n_2\coloneqq \abs{\Psi_2}^2$}{(a)--(e) and~(k)--(o)}{(f)--(j) and~(p)--(t)}}\label{fig:dyn_2_0_a12_69}
\end{figure}

An interesting phenomenon is observed in the time evolution of the dynamically unstable $\tcv{2}{0}$ vortex, as illustrated in Fig.~\ref{fig:dyn_2_0_a12_69} for $\interparam=0.7$. Solving the corresponding Bogoliubov equations yields $\max_{\qpwinding,\qpindexa}\mathrm{Im}(\omega_\qpindexa^{(\qpwinding\mkern+1mu)}/\omega_1)=\num{0.0903}$, with the maximum occurring for $\abs{\qpwinding\mkern+1mu}=2$. As expected from this result, the two-quantum vortex in component~1 first splits into two separated single-quantum vortices [Fig.~\ref{fig:dyn_2_0_a12_69}(d)] that orbit the trap counterclockwise. Surprisingly, however, the split vortices subsequently merge back together, and the entire two-component state returns approximately to its initial form [Figs.~\ref{fig:dyn_2_0_a12_69}(e) and~\ref{fig:dyn_2_0_a12_69}(j)]. This process then repeats itself with a period of $\sim\mkern-10mu \SI[mode=math]{77}{\milli\second}$ [Figs.~\ref{fig:dyn_2_0_a12_69}(k)--\ref{fig:dyn_2_0_a12_69}(t)]. A similar split-and-revival effect of a dynamically unstable multiquantum vortex was recently found for a three-quantum vortex in a three-dimensional single-component BEC~\cite{Rab2018.PRA98.023624}.

\begin{figure}[t]
\includegraphics[width=1.0\columnwidth,keepaspectratio]{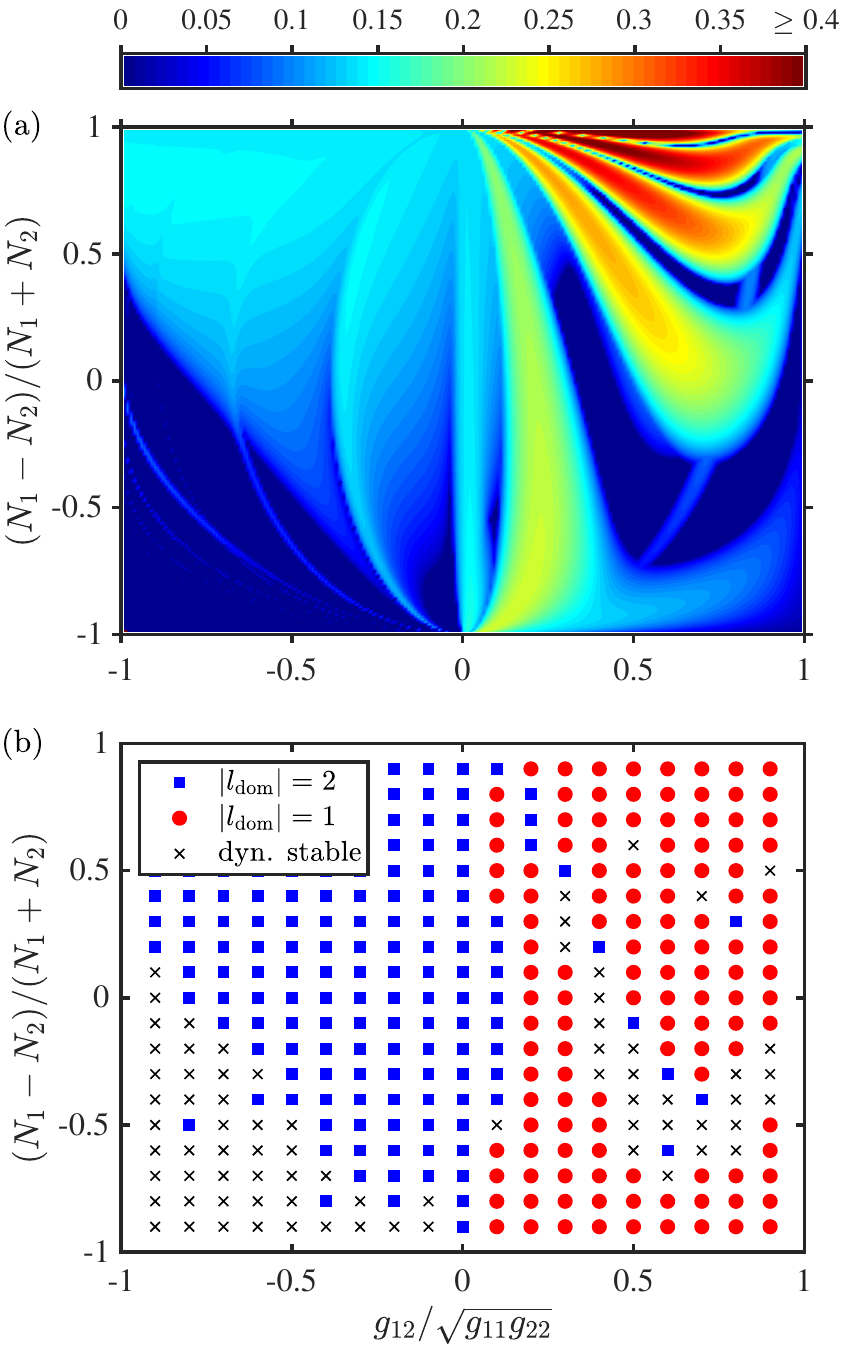}
\caption{\stabilitydiagramcaption{2}{1}{\dynstabledescription}{}}\label{fig:maxim_2_1}
\end{figure}

Moving on to the $\tcv{2}{1}$ vortex, we present its dominant dynamical instabilities in the $\interparam\mkern-1.5mu\polarization$ parameter plane in Fig.~\ref{fig:maxim_2_1}. Repulsive intercomponent interactions ($\interparam > 0$) are observed to result mainly in $\abs{\maxqpwinding}=1$, albeit interspersed with small regions of either dynamical stability or $\abs{\maxqpwinding}=2$. Attractive intercomponent interactions, on the other hand, yield only regions of dynamical stability or $\abs{\maxqpwinding}=2$. The $\tcv{2}{1}$ vortex is observed to be dynamically stable (but energetically unstable) for sufficiently strong intercomponent attraction and small polarization $\polarization$, with the latter implying that the doubly quantized vortex resides in the minority component. This stabilization corroborates the findings of Ref.~\cite{Kuo2015.PRA91.043605}, where the axisymmetric $\tcv{2}{1}$ vortex was reported to become the ground state for $\interparam \gtrapprox -1$ and $\polarization = 0$ with the addition of external rotation~\footnote{Incorporating external axial rotation by angular frequency $\Omega$ into our model would shift the excitation energies $\{\hbar\omega_\qpindexa^{(\qpwinding\mkern+1mu)}\}$ by $-\qpwinding\hbar\Omega$ and would act to lift the energetically unstable modes at $\qpwinding=-2$ and $-1$ above zero, eventually stabilizing the energetically unstable $\tcv{2}{1}$ vortices. Dynamical instabilities, in contrast, would not be suppressed by external rotation, which implies that only a dynamically stable stationary vortex state can become the rotating ground state.}. We have also carried out additional calculations indicating that the size of this stability region in the $\interparam\mkern-1.5mu\polarization$ plane tends to become larger with smaller $\geff$, as suggested in Ref.~\cite{Kuo2015.PRA91.043605}.

\begin{figure}[t]
\includegraphics[width=1.0\columnwidth,keepaspectratio]{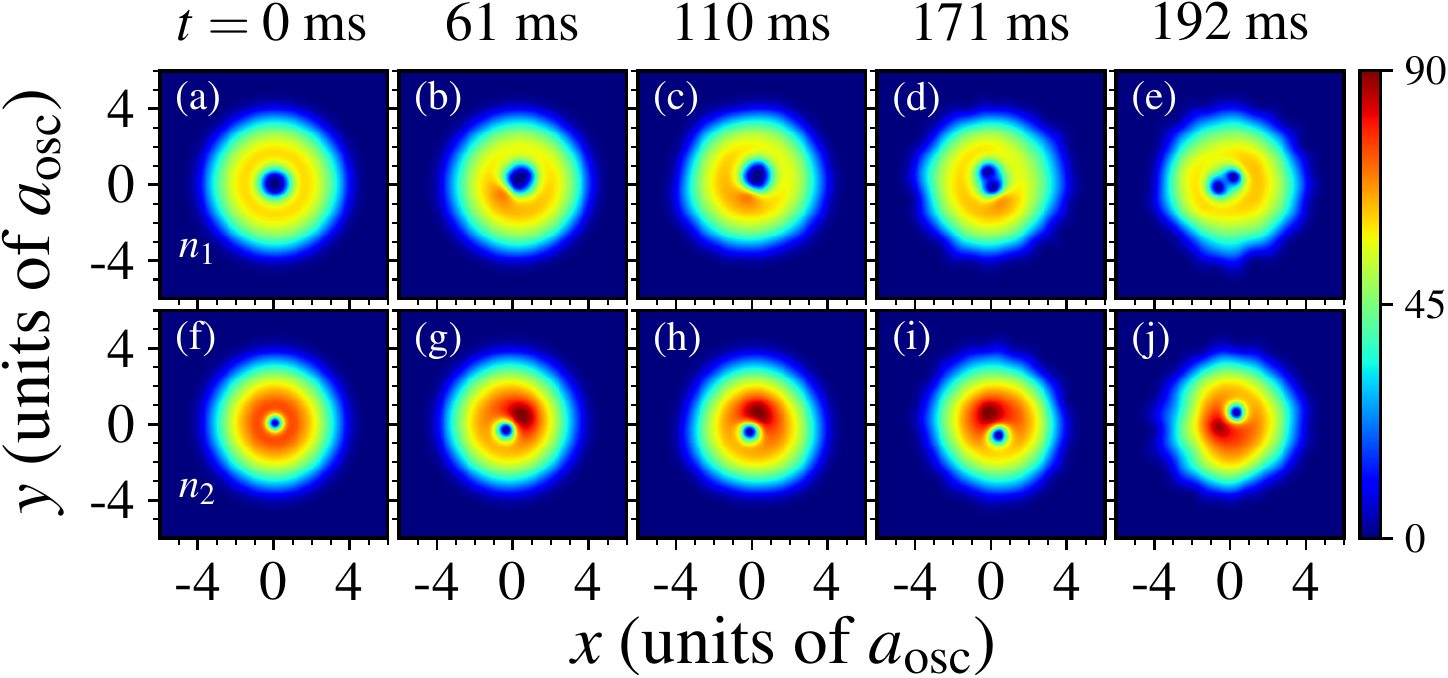}
\caption{\timeseriescaption{2}{1}{0.3}{1}{0.232}{$n_1$}{$n_2$}{(a)--(e)}{(f)--(j)}}\label{fig:dyn_2_1_a12_29.4}
\end{figure}

A representative example of the $\abs{\maxqpwinding}=1$ decay of the $\tcv{2}{1}$ vortex is shown in Fig.~\ref{fig:dyn_2_1_a12_29.4} for $\interparam = 0.3$. As indicated by the density profiles in Figs.~\ref{fig:dyn_2_1_a12_29.4}(a)--\ref{fig:dyn_2_1_a12_29.4}(c) and~\ref{fig:dyn_2_1_a12_29.4}(f)--\ref{fig:dyn_2_1_a12_29.4}(h), the two-quantum vortex in component~1 and the single-quantum vortex in component~2 first move to opposite sides of the trap center, while simultaneously orbiting it in a counterclockwise direction with a time period of $\sim\mkern-7mu \SI[mode=math]{9.9}{\milli\second}$. Subsequently, the two-quantum vortex in component~1 splits into two single-quantum vortices [Figs.~\ref{fig:dyn_2_1_a12_29.4}(d) and~\ref{fig:dyn_2_1_a12_29.4}(e)]. In our composite-vortex notation, the decay process can thus be described as a two-step splitting process, namely, the splitting of a $\tcv{2}{1}$ vortex into a $\tcv{2}{0}$ vortex and a $\tcv{0}{1}$ vortex followed by the splitting of the $\tcv{2}{0}$ vortex into two $\tcv{1}{0}$ vortices. Of these, the initial $\tcv{2}{1}$ vortex is cored, while the rest are coreless. Comparison of the density profiles at $t=\SI[mode=math]{171}{\milli\second}$ with those at $t=\SI[mode=math]{192}{\milli\second}$ reveals that all the remaining vortices continue to orbit the trap center counterclockwise.

\begin{figure}[b]
\includegraphics[width=1.0\columnwidth,keepaspectratio]{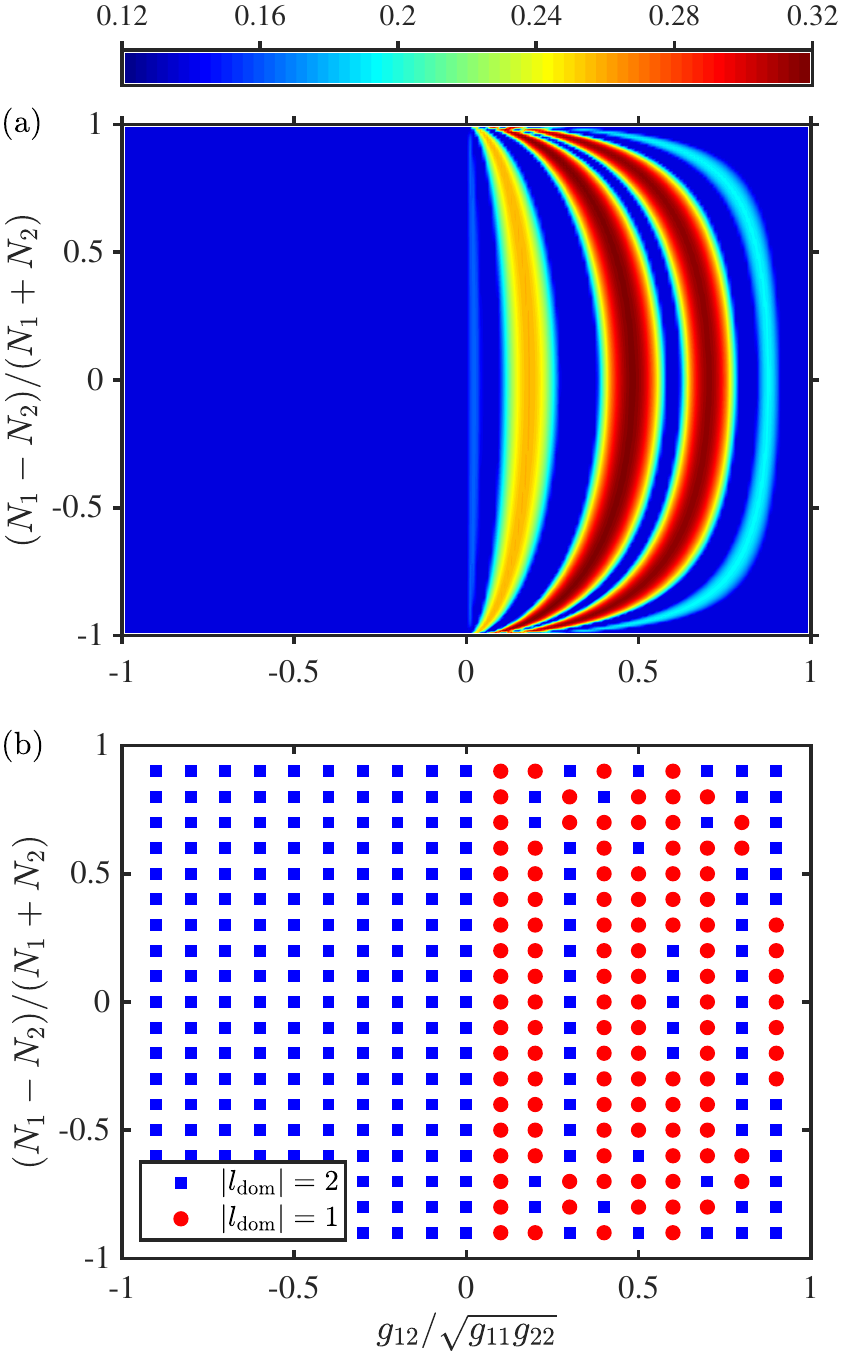}
\caption{\stabilitydiagramcaption{2}{2}{}{}}\label{fig:maxim_2_2}
\end{figure}

The behavior shown in Fig.~\ref{fig:dyn_2_1_a12_29.4} may be compared with Ref.~\cite{Tak2009.PRA79.023618}, where dynamical instabilities with $\abs{\maxqpwinding}=1$ were found for $\thcv{0}{1}{2}$ vortices in three-component spinor BECs with antiferromagnetic spin--spin interactions. Those instabilities, however, were not associated with splitting of the vortices but with segregation of the different spin components. Although the principal effect of the $\abs{\qpwinding\mkern+1mu}=1$ instabilities in our two-component system is to split the composite vortex, the occurrence of $\abs{\maxqpwinding}=1$ only for $\interparam > 0$ in Fig.~\ref{fig:maxim_2_1} suggests that the segregation tendency due to repulsive intercomponent interactions plays a role in amplifying these $\abs{\qpwinding\mkern+1mu}=1$ instabilities. Our results on single-quantum composite vortices (Appendix~\ref{app:single-quantum_results}) also support this inference.

The $\abs{\maxqpwinding}$ diagram of the $\tcv{2}{-1}$ vortex is fairly similar to that of the $\tcv{2}{1}$ vortex in that they both contain regions of dynamical stability, regions of $\abs{\maxqpwinding}=1$, and regions of $\abs{\maxqpwinding}=2$. There are two main differences: (i)~the $\tcv{2}{-1}$ vortex can have $\abs{\maxqpwinding} =1$ for both $\interparam < 0$ and $\interparam > 0$, whereas for the $\tcv{2}{1}$ vortex this value is exclusive to $\interparam > 0$; (ii)~the $\tcv{2}{-1}$ vortex does not exhibit dynamical stability for $\interparam > 0$. The stability diagrams of the $\tcv{2}{-1}$ vortex, along with a representative time series, can be found in Appendix~\ref{app:+2_-1}.

\begin{figure}[t]
\includegraphics[width=1.0\columnwidth,keepaspectratio]{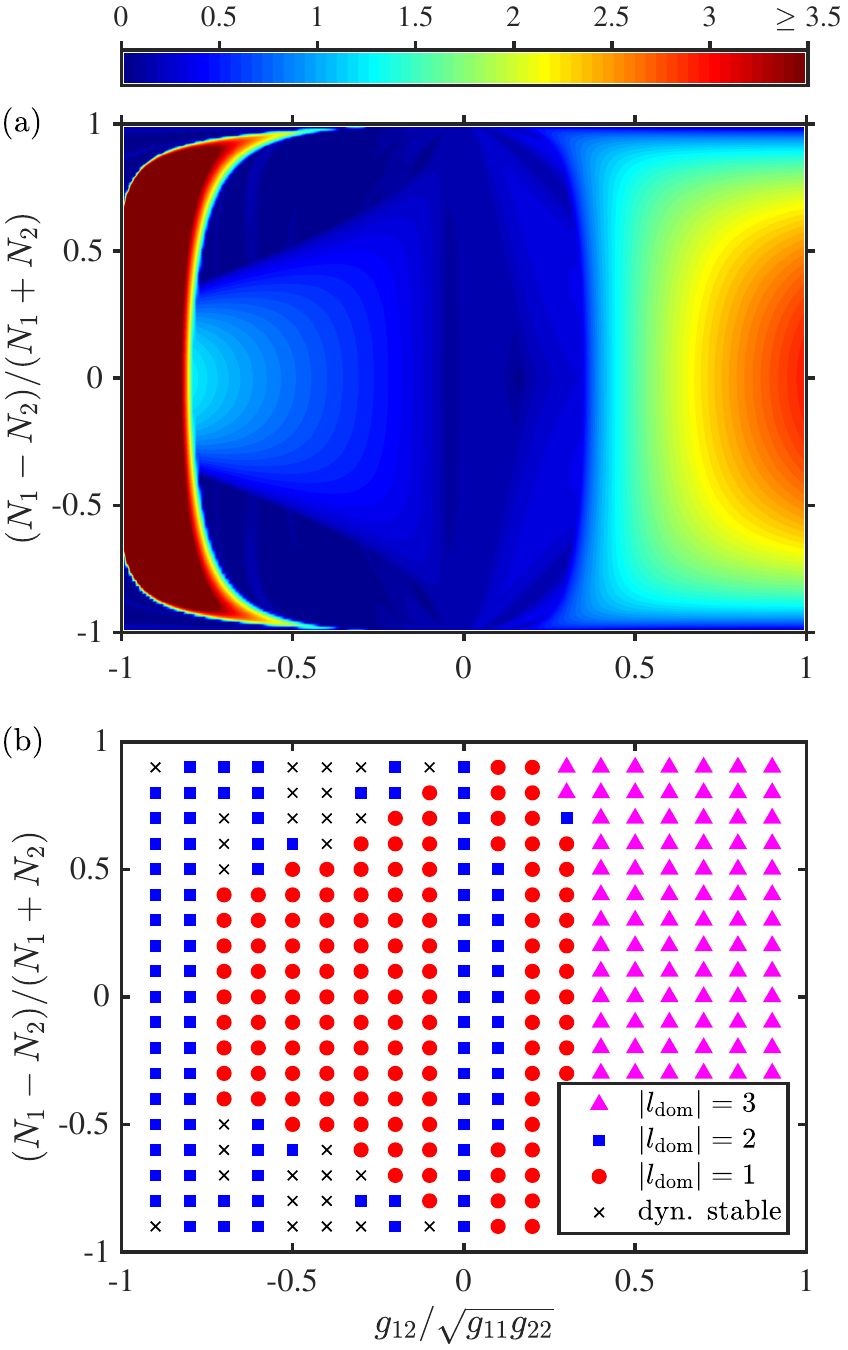}
\caption{\stabilitydiagramcaption{2}{-2}{\dynstabledescription}{}}\label{fig:maxim_2_-2}
\end{figure}

A common feature of the $\tcv{2}{0}$ and $\tcv{2}{\pm 1}$ vortices is that their instabilities occur only for $\abs{\qpwinding\mkern+1mu}= 1$ and $2$. The $\tcv{2}{\pm 2}$ vortices, in contrast, may exhibit energetic and dynamical instabilities for $\abs{\qpwinding\mkern+1mu}=3$ as well. For the $\tcv{2}{2}$ vortex, these are typically energetic instabilities with $\qpwinding=-3$, which develop into dynamical instabilities only in narrow regions of the $\interparam\mkern-1.5mu\polarization$ space and are so weak that they are dominated by much stronger coexisting instabilities with $\abs{\qpwinding\mkern+1mu}=1$ or $2$. Accordingly, the $\abs{\maxqpwinding}$ diagram of the $\tcv{2}{2}$ vortex [Fig.~\ref{fig:maxim_2_2}(b)] exhibits only the values $1$ and $2$; note that the $\tcv{2}{2}$ vortex is dynamically unstable for all values of $\interparam$ and $\polarization$. Figure~\ref{fig:maxim_2_2}(b) also indicates that the dominant dynamical instability of the $\tcv{2}{2}$ vortex is provided exclusively by the $\abs{\qpwinding\mkern+1mu}= 2$ modes for $\interparam\leq 0$, whereas the region $\interparam > 0$ corresponds predominantly to $\abs{\maxqpwinding}=1$. 

The stability diagrams of the $\tcv{2}{-2}$ vortex are shown in Fig.~\ref{fig:maxim_2_-2}. For $\interparam<0$, we observe regions of dynamical instability with $\abs{\maxqpwinding}=1$ and $2$, as well as small regions where the $\tcv{2}{-2}$ vortex is dynamically stable (albeit energetically unstable). The region $\interparam \geq 0.4$, on the other hand, corresponds exclusively to the exotic instability mode $\abs{\maxqpwinding}=3$, which does not occur for the two-quantum vortex in single-component BECs---more generally, a $\windingnumber$-quantum vortex in a harmonically trapped single-component BEC can only exhibit dynamical instabilities with $\abs{\qpwinding\mkern+1mu} \leq \abs{\windingnumber}$. Due to the prominence of $\abs{\maxqpwinding}=1$ for $\interparam < 0$, Fig.~\ref{fig:maxim_2_-2}(b) does not fully corroborate the observation made in Ref.~\cite{Ish2013.PRA88.063617} that, for $\tcv{\windingnumber}{-\windingnumber}$ vortices, $\abs{\maxqpwinding}$ tends to be even for $\interparam < 0$ and odd for $\interparam > 0$.

\begin{figure}[t]
\includegraphics[width=1.0\columnwidth,keepaspectratio]{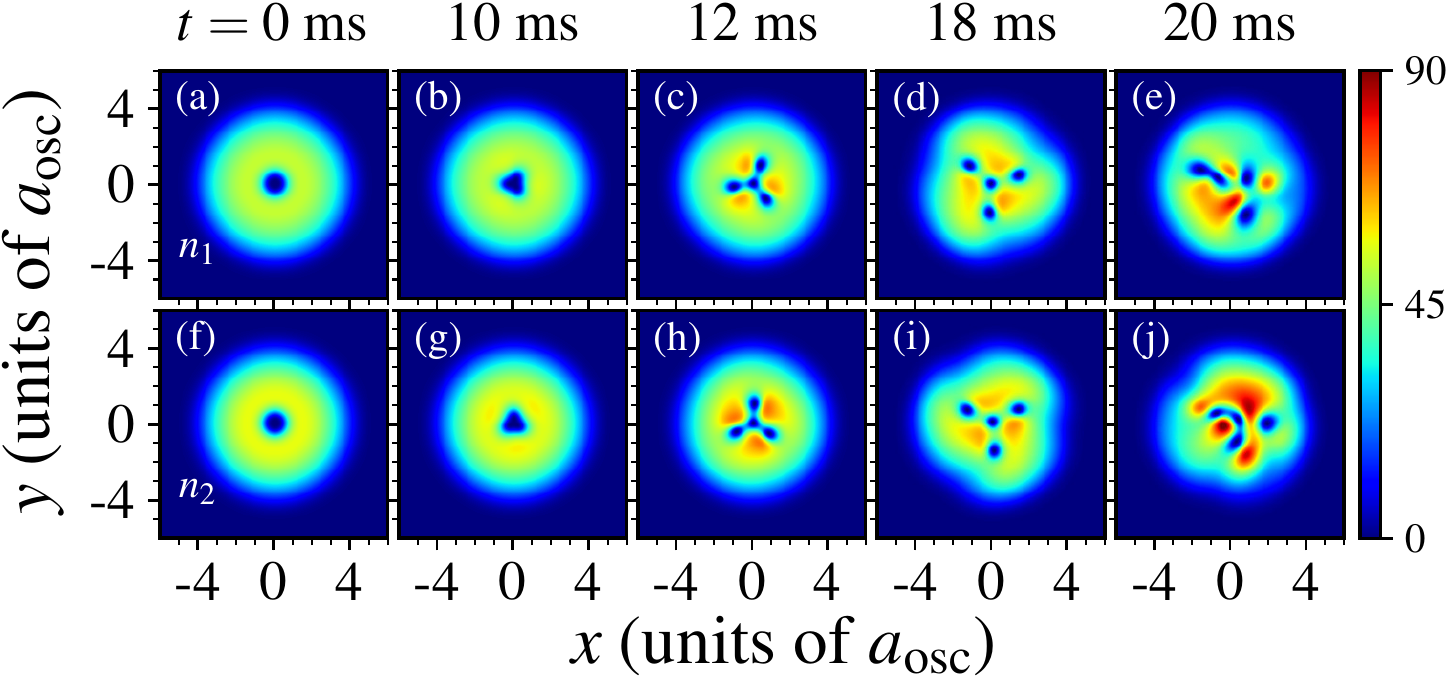}
\caption{\timeseriescaption{2}{-2}{0.8}{3}{1.41}{$n_1$}{$n_2$}{(a)--(e)}{(f)--(j)}}\label{fig:dyn_2_-2_a12_78.3}
\end{figure}

The full time evolution corresponding to the dominant $\abs{\qpwinding\mkern+1mu}=3$ instability of the $\tcv{2}{-2}$ vortex is illustrated for $\interparam=0.8$ in Fig.~\ref{fig:dyn_2_-2_a12_78.3}. During the first \SI[mode=text]{18}{\milli\second} of the evolution, the doubly charged vortex in each component splits into four single-quantum vortices in a threefold-symmetric pattern. In component~1, the three outward-moving vortices have the winding number $+1$ and the central one has $-1$; in component~2, the winding numbers are the opposite of these. In the composite-vortex notation, this translates to the splitting of the $\tcv{2}{-2}$ vortex into one $\tcv{-1}{1}$ vortex, three $\tcv{1}{0}$ vortices, and three $\tcv{0}{-1}$ vortices; here the $\tcv{2}{-2}$ vortex and the $\tcv{-1}{1}$ vortex are cored, while the other six are coreless. Concurrently with their radial motion, the off-axis vortices orbit the center in a counterclockwise direction in component~1 and in a clockwise direction in component~2. At later times, the shapes of both the individual vortex cores and the overall condensate densities $n_1$ and $n_2$ become highly irregular [Figs.~\ref{fig:dyn_2_-2_a12_78.3}(e) and~\ref{fig:dyn_2_-2_a12_78.3}(j)].

\section{\label{sec:discussion}Conclusions}

In summary, we have studied the dynamical instabilities and splitting of axisymmetric composite vortices in harmonically trapped two-component BECs using a two-dimensional GP model. Limiting the paper to singly and doubly quantized composite vortices in the miscible regime $g_{12}^2 < \intrag{1}\intrag{2}$, we formed the vortex stability diagrams in the two-dimensional parameter space consisting of the relative intercomponent interaction strength $\interparam \coloneqq g_{12}/\sqrt{\intrag{1}\intrag{2}}$ and the particle-number polarization $\polarization\coloneqq\left(N_1-N_2\right)/\left(N_1+N_2\right)$. To wit, for each winding-number pair $\tcv{\windingnumber_1}{\windingnumber_2}$ and parameter point $\pointcoord{\interparam}{\polarization}$, where $2 \geq \windingnumber_1 \geq \abs{\windingnumber_2}$ and $\interparam$ and $\polarization$ take values from $-1$ to $1$, we solved the Bogoliubov equations to obtain the magnitude and multipolarity of the dominant dynamical instability of that stationary vortex state. The decay behaviors associated with different multipolarities $\abs{\maxqpwinding}$ were then demonstrated by solving the full time evolution for slightly perturbed versions of the stationary vortices from the time-dependent GP equations. 

Several decay modes stemming from the multicomponent nature of the system, and thus absent for vortex states in scalar BECs, were discovered. The cored single-quantum composite vortices $\tcv{\windingnumber_1}{\windingnumber_2}=\tcv{1}{\pm 1}$ exhibited regions of dynamical instability with $\abs{\maxqpwinding}=1$; this is to be contrasted with the axisymmetric single-quantum vortex state in a scalar BEC, which is always dynamically stable. The decay of the unstable $\tcv{1}{\pm 1}$ vortices was found to involve opposite displacements of the constituent vortex cores from the trap center and a significant reduction in the density overlap $\iint \abs{\Psi_1\Psi_2}\, \intmeasure$. Although such behavior suggests a connection between these dynamical instabilities and the phase-separation tendency associated with the intercomponent repulsion ($\interparam > 0$), we note that the $\tcv{1}{-1}$ vortex had strong dynamical instabilities also for $\interparam < 0$. The coreless $\tcv{1}{0}$ vortex turned out to be always dynamically stable.

The two-quantum composite vortices $\tcv{2}{0}$, $\tcv{2}{\pm 1}$, and $\tcv{2}{\pm 2}$ were found to exhibit splitting behavior that has no counterpart for the two-quantum vortex in a single-component BEC. The $\abs{\maxqpwinding}=1$ instabilities of the $\tcv{2}{\pm 1}$ and $\tcv{2}{\pm 2}$ vortices induced a two-step splitting process in which the constituent two-quantum vortices drifted away from the trap center before splitting individually; the $\abs{\maxqpwinding}=2$ instabilities, in contrast, resulted in the splitting of the vortices directly at the center. For the $\tcv{2}{0}$ and $\tcv{2}{2}$ vortices, we demonstrated a split-and-revival phenomenon, in which the vortex returned close to its initial form after splitting temporarily. For the $\tcv{2}{-2}$ vortex, we observed a threefold-symmetric splitting pattern in which each doubly quantized vortex split into three singly quantized vortices of like sign and one of unlike sign. We showed that this splitting instability, which was discovered in Ref.~\cite{Ish2013.PRA88.063617} and attributed to superfluid--superfluid counterflow~\cite{Tak2010.PRL105.205301,Ham2011.PRL106.065302,Ish2011.PRA83.063602}, dominates over other instabilities of the $\tcv{2}{-2}$ vortex in a large region of the $\interparam\mkern-1.5mu\polarization$ space. The threefold splitting pattern, which does not occur for the single-component two-quantum vortex, should therefore be amenable to experimental verification as long as the $\tcv{2}{-2}$ vortex can be realized. This should be achievable by applying the topological phase engineering technique~\cite{Lea2002.PRL89.190403,Lea2003.PRL90.140403,Shi2004.PRL93.160406,Kum2006.LP16.371,Kum2006.PRA73.063605,Iso2007.PRL99.200403,Kuw2010.JPSJ79.034004,Shi2011.JPhysB44.075302} to a two-component BEC composed of two hyperfine spin states with opposite values of $g_F m_F$, where $g_F$ is the Land\'{e} factor. In addition, more highly charged counter-rotating vortex states $\tcv{\windingnumber}{-\windingnumber}$, where $\windingnumber$ is an even number $\geq 4$, could potentially be created by applying the vortex pump~\cite{Mot2007.PRL99.250406,Xu2008.PRA78.043606,Xu2008.NJP11.055019,Xu2010.PRA81.053619,Kuo2010.JLTP161.561,Kuo2013.PRA87.033623} to such a condensate.

\begin{figure}[t]
\includegraphics[width=1.0\columnwidth,keepaspectratio]{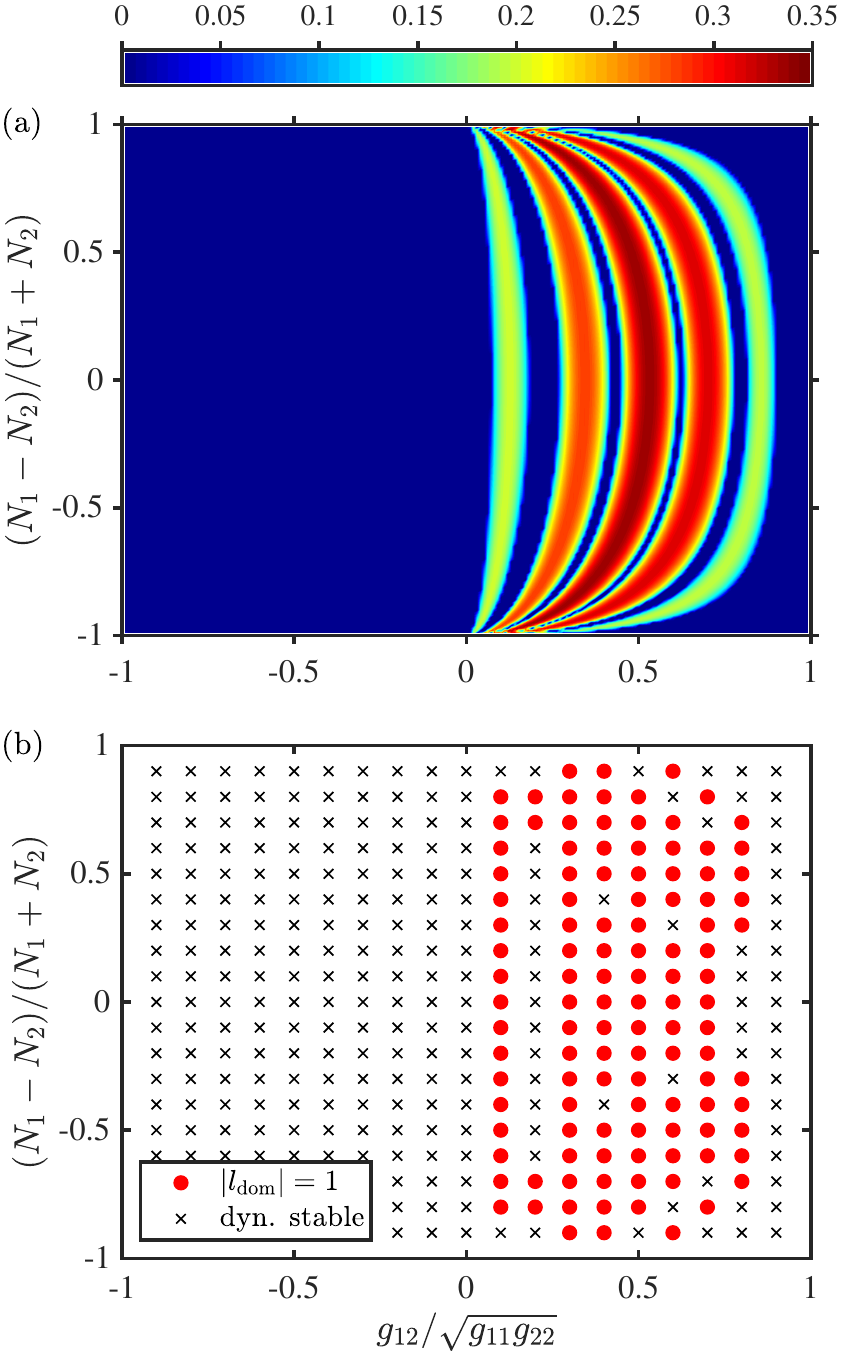}
\caption{\stabilitydiagramcaption{1}{1}{\dynstabledescription}{}}\label{fig:maxim_1_1}
\end{figure}

Besides increasing $\max_j \abs{\windingnumber_j}$ above $2$ to reveal further splitting patterns~\footnote{Reference~\cite{Ish2013.PRA88.063617} addresses $\tcv{3}{-3}$ and $\tcv{10}{-10}$ vortices at $\pointcoord{\interparam}{\polarization}=\pointcoord{0.9}{0}$ and reports the dominant instability modes $\abs{\maxqpwinding} = 5$ and $15$, respectively.}, there are several other ways to extend this paper. Incorporating the immiscible regime $\interparam > 1$ and the associated nonaxisymmetric stationary vortex states would present an opportunity to investigate the interplay~\cite{Ban2017.PRA96.043603} between vortex dynamics, phase separation~\cite{Nav2009.PRA80.023613,Lee2016.PRA94.013602,Mis2018.NJP20.043052}, and fluid interface instabilities~\cite{Sas2009.PRA80.063611,Kob2011.PRA83.043623,Kad2012.PRA85.013602}. Generalization to three dimensions would enable us to study the splitting-induced intertwining of vortices~\cite{Mot2003.PRA68.023611,Huh2006.PRL97.110406,Mat2006.PRL97.180409} as well as the dynamics of composite defects consisting of vortex lines and interfaces~\cite{Tak2006.JPSJ75.063601,Kas2013.PRA88.013620,Kas2013.JPCM25.404213,Kan2019.PRL122.095301}. It would also be interesting to examine finite-temperature effects, given that the presence of the thermal component is predicted to stabilize an axisymmetric vortex energetically~\cite{Iso1999.PRA59.2203,Vir2001.PRL86.2704,Vir2001.JPCM13.L819}.

Finally, as a cautionary remark, we note that the mean-field GP equations~\eqref{eq:2dgpe} may not provide a complete picture of the two-component BEC in the vicinity of the critical value $\interparam=-1$, where they predict the condensate mixture to collapse. The concern here is that the first beyond-mean-field correction to the energy, the so-called Lee--Huang--Yang~(LHY) term~\cite{Lee1957.PR106.1135,Lar1963.AnnPhys24.89}, may no longer be negligible in this regime. In fact, recent theoretical~\cite{Pet2015.PRL115.155302,Pet2016.PRL117.100401} and experimental~\cite{Cab2018.Sci359.301,Sem2018.PRL120.235301} works have demonstrated that the LHY correction can stabilize the two-component BEC against the collapse and lead to the formation of self-bound quantum droplets for $\interparam < -1$. Although the incorporation of the LHY term into the two-component model would likely alter the features next to $\interparam = -1$ in some of the vortex stability diagrams, it would not change our present conclusions, all of which have been drawn independently of the behavior near this $\interparam$ value. For an account of the stability of self-bound composite vortices for $\interparam < -1$, we refer to the recent work of Kartashov, Malomed, Tarruell, and Torner~\cite{Kar2018.PRA98.013612}.

\begin{acknowledgments}
P.K. acknowledges funding from the Technology Industries of Finland Centennial Foundation and the Academy of Finland (Grant No.~308632). A.R. acknowledges partial support from Provincia Autonoma di Trento for this paper. S.B. and D.A. acknowledge the use of the Vikram-100 high-performance computing cluster at the Physical Research Laboratory, Ahmedabad, for some of the simulations presented here. The authors thank Rukmani Bai, Boris Malomed, Sukla Pal, Tapio Simula, and Kuldeep Suthar for insightful discussions.
\end{acknowledgments}

\appendix

\section{\label{app:single-quantum_results}Decay of \texorpdfstring{{\boldmath{$\tcv{1}{\pm 1}$}}}{(1,1) and (1,-1)} vortices}

This appendix presents the stability diagrams and representative decay dynamics of $\tcv{1}{1}$ and $\tcv{1}{-1}$ vortices. The main observations from these calculations were outlined in Sec.~\ref{subsec:single-quantum_results}. We omit the data for the $\tcv{1}{0}$ vortex, because this state is always dynamically stable in our parameter set~$\parameterset{1}$.

\subsection{\texorpdfstring{{\boldmath{$\tcv{1}{1}$}}}{(1,1)} vortex}

The stability diagrams of the $\tcv{1}{1}$ vortex are shown in Fig.~\ref{fig:maxim_1_1}. The vortex state is observed to be dynamically unstable with $\abs{\maxqpwinding}=1$ over most of the region $\interparam > 0$, whereas for $\interparam \leq 0$ it is only energetically unstable.

\begin{figure}[t]
\includegraphics[width=1.0\columnwidth,keepaspectratio]{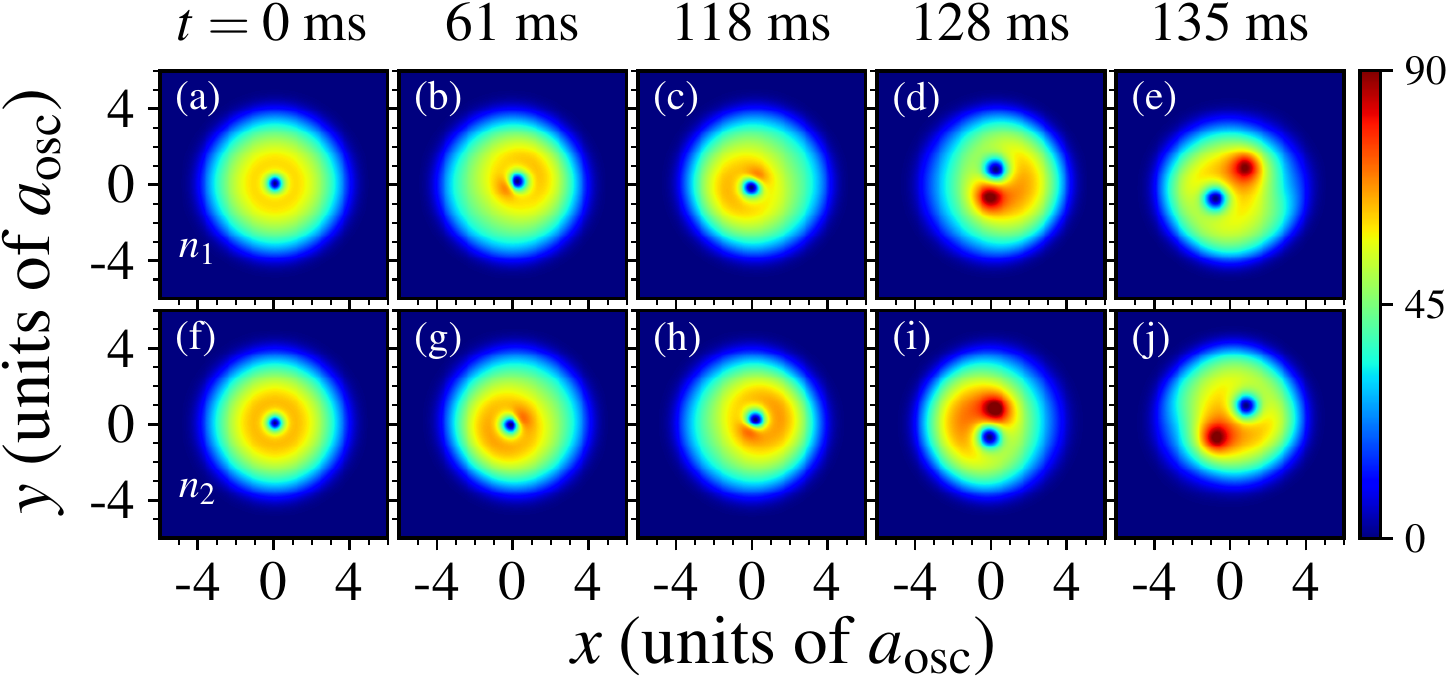}
\caption{\timeseriescaption{1}{1}{0.6}{1}{0.231}{$n_1$}{$n_2$}{(a)--(e)}{(f)--(j)}}\label{fig:dyn_1_1_a12_58.7}
\end{figure}

\begin{figure}[b]
\includegraphics[width=1.0\columnwidth,keepaspectratio]{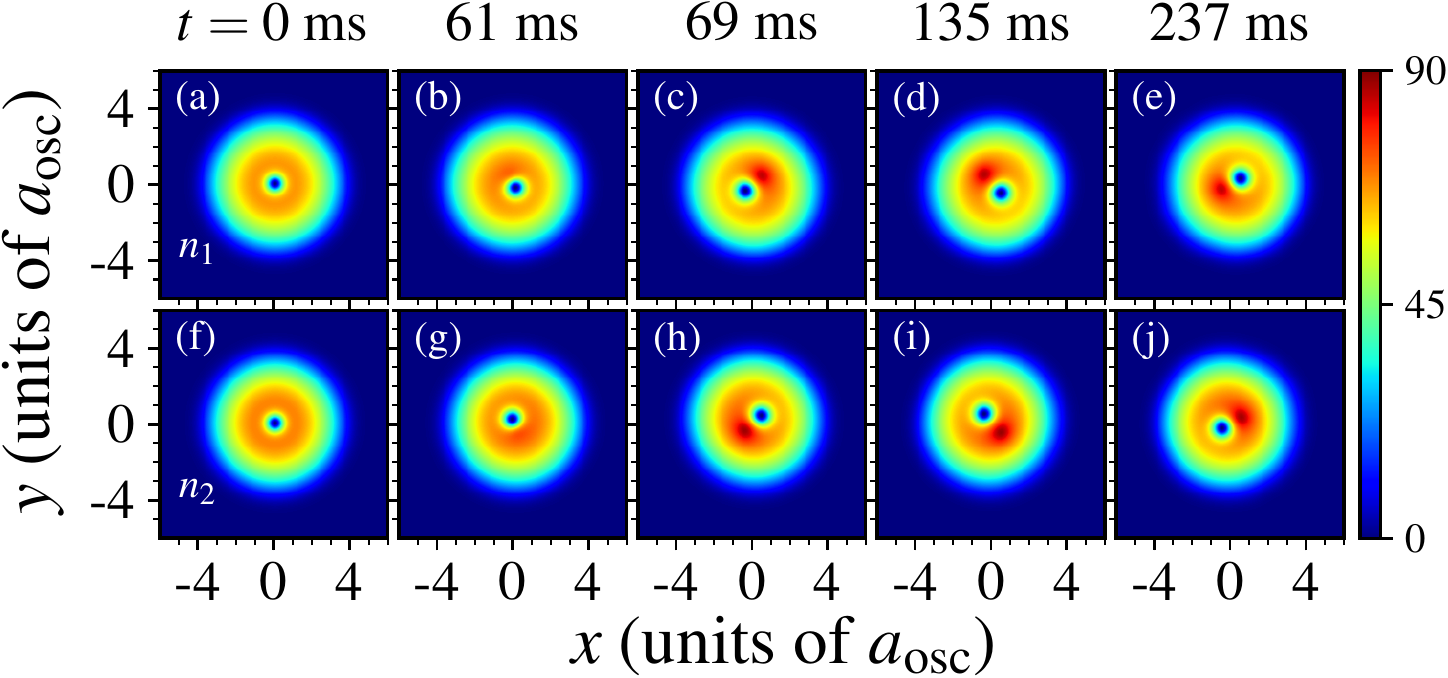}
\caption{\timeseriescaption{1}{1}{0.3}{1}{0.240}{$n_1$}{$n_2$}{(a)--(e)}{(f)--(j)}}\label{fig:dyn_1_1_a12_29.4}
\end{figure}

To investigate the decay dynamics associated with the $\abs{\maxqpwinding}=1$ stripes in Fig.~\ref{fig:maxim_1_1}, we first consider the case $\interparam=\num{0.6}$, which in set~$\parameterset{2}$ corresponds to $a_{12} = \num{58.7}\,\bohrradius$. The time evolution of this vortex state after a small perturbation is illustrated in Fig.~\ref{fig:dyn_1_1_a12_58.7}. The instability is observed to correspond to dynamics where the vortices are displaced in opposite directions from the trap center. Both vortices orbit the center counterclockwise, and the separation between them is found to oscillate with a period of $\sim\mkern-4mu \SI[mode=math]{25}{\milli\second}$. This oscillatory behavior is to be contrasted with the case shown in Fig.~\ref{fig:dyn_1_1_a12_29.4}, where $\interparam=0.3$. In the latter case, the separation of the two vortex cores remains constant for $t \geq \SI[mode=math]{69}{\milli\second}$; in other words, the vortex pair is observed to rotate as a rigid body after their initial separation stage.

\subsection{\texorpdfstring{{\boldmath{$\tcv{1}{-1}$}}}{(1,-1)} vortex}

\begin{figure}[t]
\includegraphics[width=1.0\columnwidth,keepaspectratio]{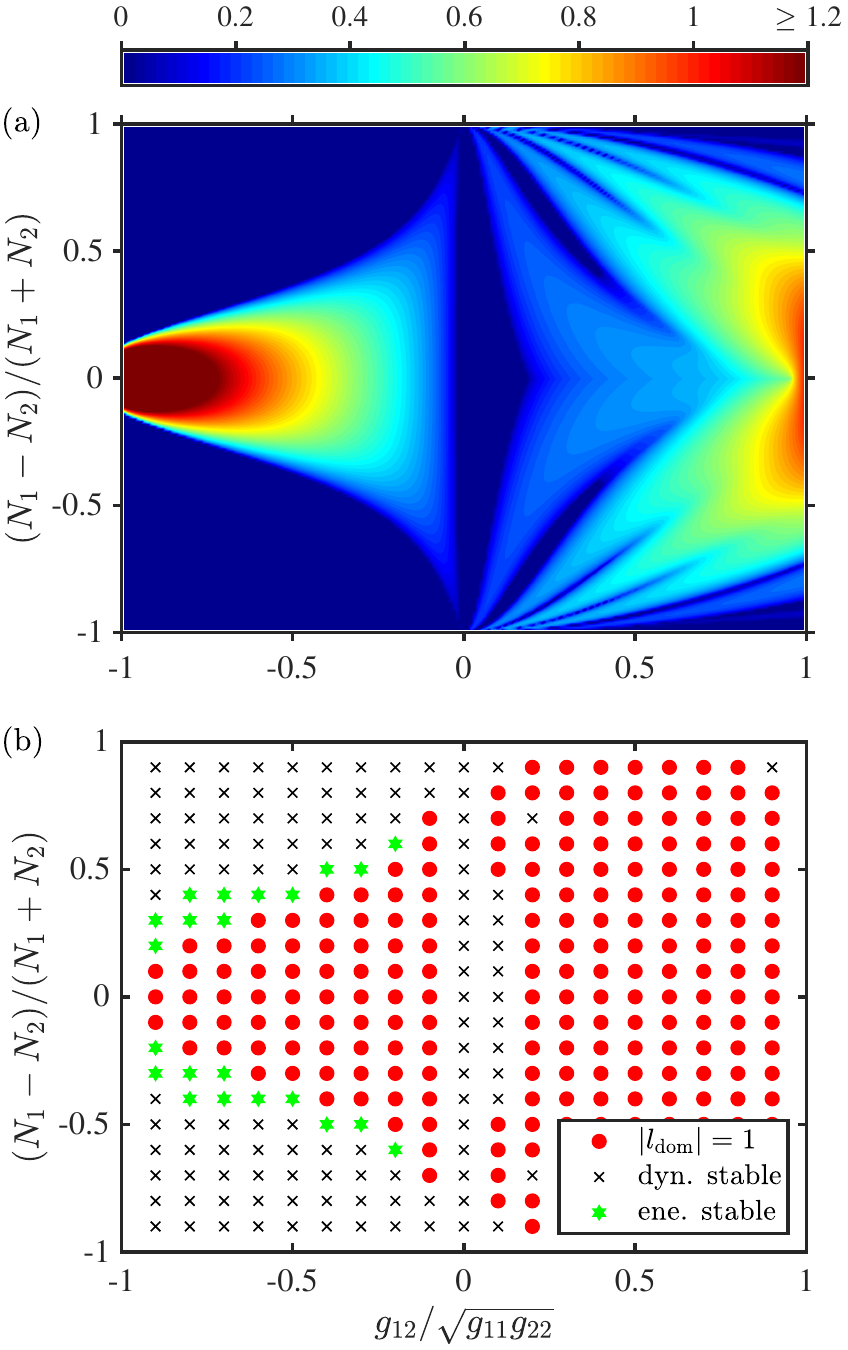}
\caption{\stabilitydiagramcaption{1}{-1}{\dynstabledescription}{\enestabledescription}}\label{fig:maxim_1_-1}
\end{figure}

Figure~\ref{fig:maxim_1_-1} presents the stability diagrams of the singly quantized counter-rotating vortex, i.e., the $\tcv{1}{-1}$ vortex. The region corresponding to $\abs{\maxqpwinding}=1$ covers most of the $\interparam\mkern-1.5mu\polarization$ space. Interestingly, small regions of energetic stability exist for $\interparam < 0$; it should be noted that this \emph{local} energetic stability does not mean that the $\tcv{1}{-1}$ vortex has lower total energy than the vortex-free state, which remains the only \emph{globally} stable solution of Eqs.~\eqref{eq:axisymGPE}.

\begin{figure}[t]
\includegraphics[width=1.0\columnwidth,keepaspectratio]{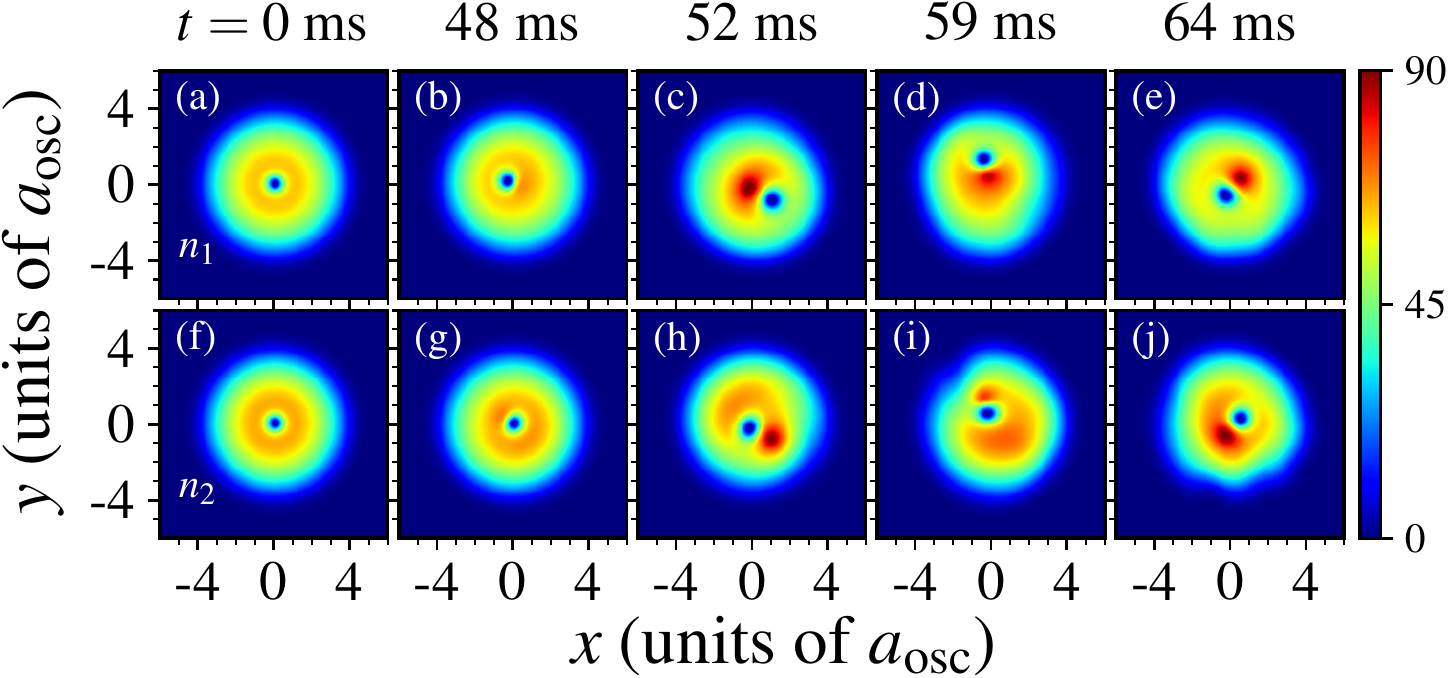}
\caption{\timeseriescaption{1}{-1}{0.5}{1}{0.283}{$n_1$}{$n_2$}{(a)--(e)}{(f)--(j)}}\label{fig:dyn_1_-1_a12_48.9}
\end{figure}

The typical decay of the $\tcv{1}{-1}$ vortex is illustrated in Fig.~\ref{fig:dyn_1_-1_a12_48.9} for $\interparam=\num{0.5}$, which corresponds to $a_{12} = \num{48.9}\,\bohrradius$ in set~$\parameterset{2}$. Inspection of Fig.~\ref{fig:dyn_1_-1_a12_48.9} and the associated video~\cite{supplemental_videos} reveals that, from $t \approx \SI[mode=math]{48}{\milli\second}$ onward, the two oppositely charged vortices drift away from the trap center and begin to orbit it in opposite directions: counterclockwise in component~1 and clockwise in component~2.

It is interesting to note that Brtka, Gammal, and Malomed~\cite{Brt2010.PRA82.053610} studied the stability of the $\tcv{1}{-1}$ vortex in the case of attractive intracomponent interactions ($\intrag{1}=\intrag{2}<0$) and found dynamical instabilities for values of $\abs{\qpwinding\mkern+1mu}$ from 1 up to 4, as well as regions of dynamical stability (energetic stability was not examined). The higher-symmetry instability modes, which have $\abs{\qpwinding\mkern+1mu}\geq 2$ and are absent in our self-repulsive system, were different from the $\abs{\qpwinding\mkern+1mu}=1$ mode in that they did not lead to appreciable motion of the vortex cores, but to partitioning of the surrounding condensates.

\section{\label{app:two-quantum_results}Decay of \texorpdfstring{{\boldmath{$\tcv{2}{\pm 1}$}} and {\boldmath{$\tcv{2}{\pm 2}$}}}{(2,1), (2,-1), (2,2), and (2,-2)} vortices}

This appendix provides additional numerical results on the dynamical instabilities and splitting of two-quantum composite vortices, complementary to the main findings already discussed in Sec.~\ref{subsec:two-quantum_results}.

\subsection{\texorpdfstring{{\boldmath{$\tcv{2}{1}$}}}{(2,1)} vortex}

\begin{figure}[t]
\includegraphics[width=0.91\columnwidth,keepaspectratio]{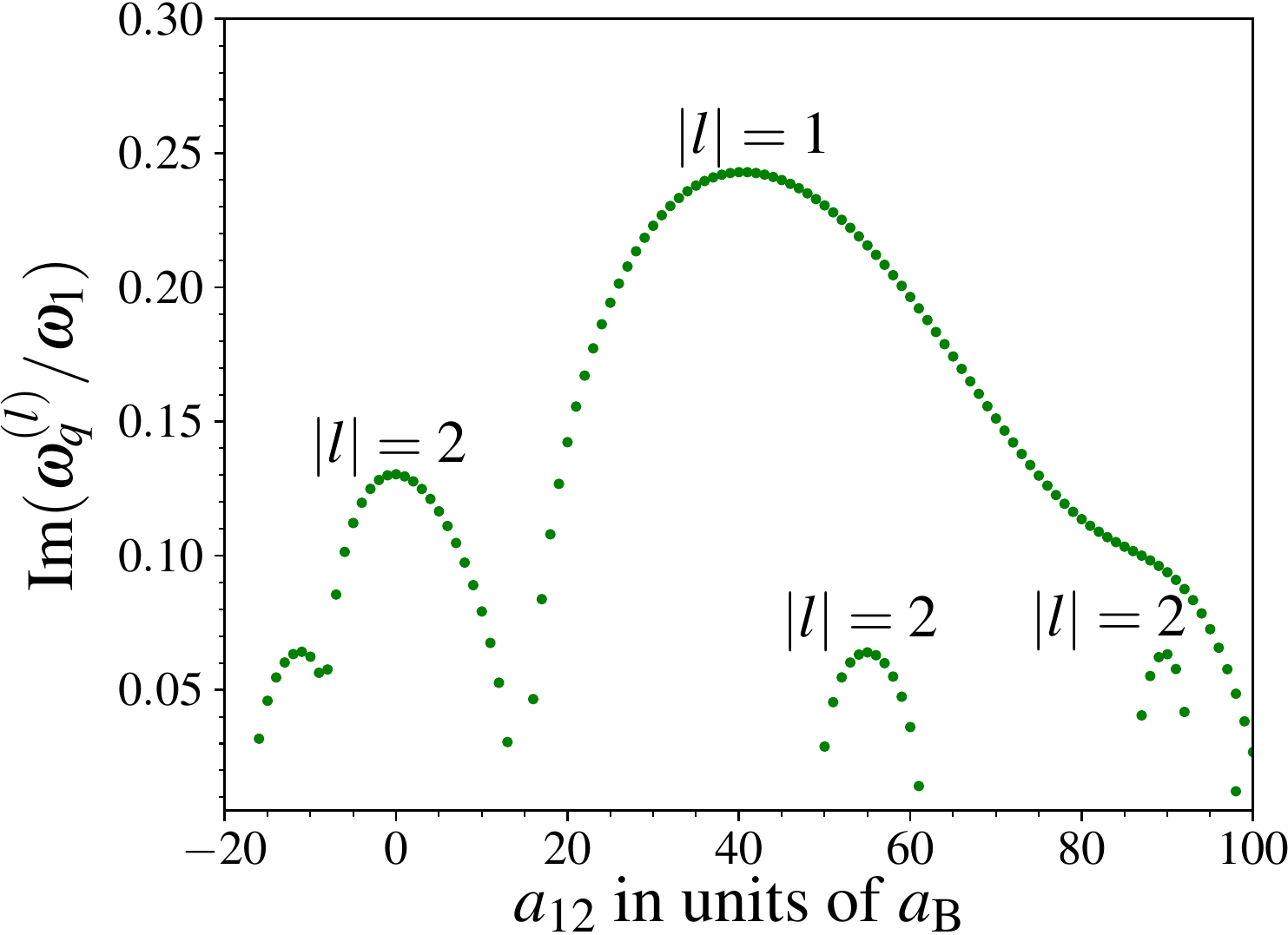}
\caption{\imaginarypartscaption{2}{1}}\label{fig:im_mode_evo_2_1}
\end{figure}

\begin{figure}[b]
\includegraphics[width=1.0\columnwidth,keepaspectratio]{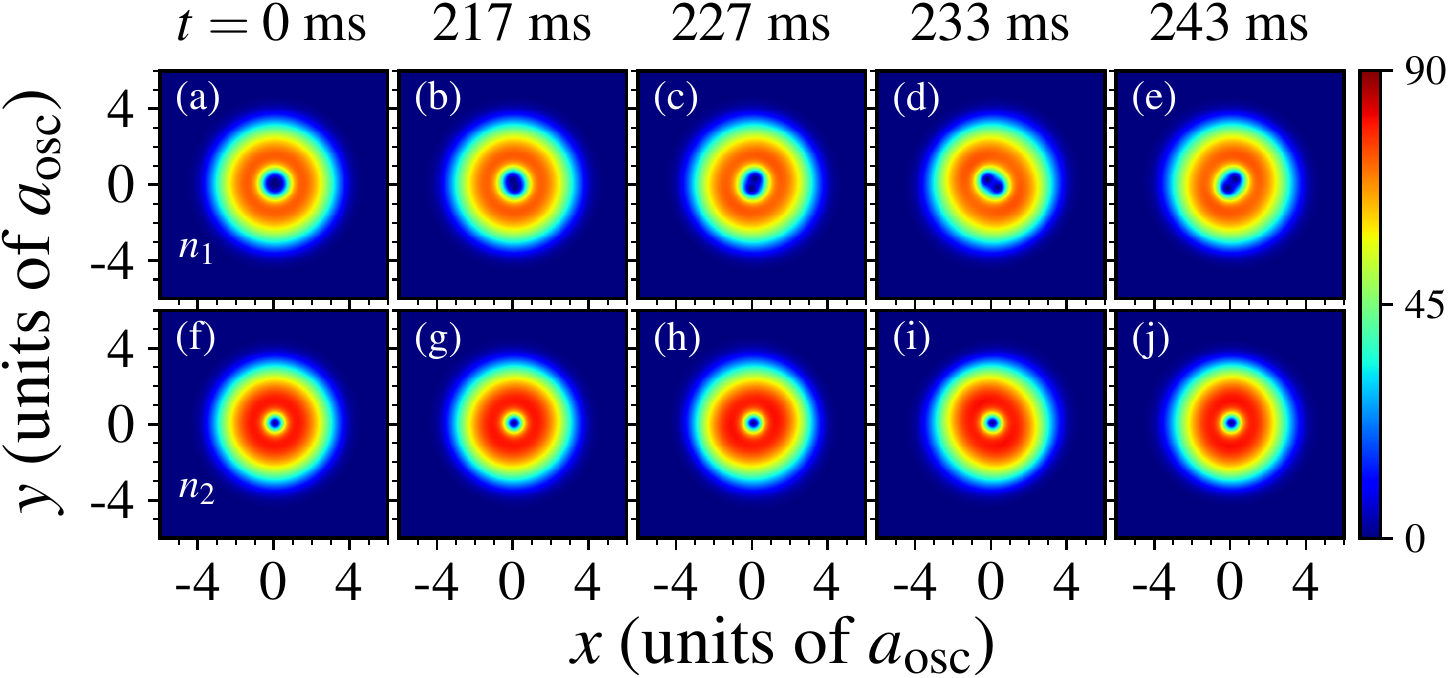}
\caption{\timeseriescaption{2}{1}{-0.1}{2}{0.0552}{$n_1$}{$n_2$}{(a)--(e)}{(f)--(j)}}\label{fig:dyn_2_1_a12_-9.8}
\end{figure}

For the $\tcv{2}{1}$ vortex, we plot the imaginary parts of the excitation frequencies as a function of $a_{12}$ in Fig.~\ref{fig:im_mode_evo_2_1}. From the figure, we can infer that the vortex is dynamically unstable in the entire range shown, $\num{-20}\,\bohrradius\lesssim a_{12}\lesssim \num{100}\,\bohrradius$, since there are excitation frequencies with positive imaginary parts at all values of $a_{12}$. In terms of the dimensionless interaction parameter $\interparam$, this range of $a_{12}$ corresponds to $\num{-0.2} \lesssim\interparam\lesssim \num{1.0}$. For $\interparam\gtrsim \num{0.13}$, the dynamical instability arises predominantly from excitations with $\abs{\qpwinding\mkern+1mu} = 1$ and results in the dynamics illustrated in Fig.~\ref{fig:dyn_2_1_a12_29.4} and discussed in Sec.~\ref{subsec:two-quantum_results}.

On the other hand, for $\interparam\lesssim 0.13$ the dominant contribution to the instability comes from the $\abs{\qpwinding\mkern+1mu} = 2$ modes. We illustrate this regime in Fig.~\ref{fig:dyn_2_1_a12_-9.8} with the case $\interparam= \num{-0.1}$, i.e., $a_{12} = \num{-9.8}\,\bohrradius$. At $t \approx \SI[mode=math]{220}{\milli\second}$, the doubly charged vortex in component~1 splits into two singly charged vortices, which orbit the trap center in a counterclockwise direction. Contrary to the $\abs{\qpwinding\mkern+1mu}=1$ dynamics in Fig.~\ref{fig:dyn_2_1_a12_29.4}, the doubly charged vortex is not expelled from the center before it splits; the singly charged vortex in component~2 also remains at the center throughout the evolution [Figs.~\ref{fig:dyn_2_1_a12_-9.8}(f)--\ref{fig:dyn_2_1_a12_-9.8}(j)]. In the composite-vortex notation, the process shown in Fig.~\ref{fig:dyn_2_1_a12_-9.8} can be expressed as the splitting of the $\tcv{2}{1}$ vortex into two $\tcv{1}{0}$ vortices and one $\tcv{0}{1}$ vortex.

\subsection{\label{app:+2_-1}\texorpdfstring{{\boldmath{$\tcv{2}{-1}$}}}{(2,-1)} vortex}

\begin{figure}[t]
\includegraphics[width=1.0\columnwidth,keepaspectratio]{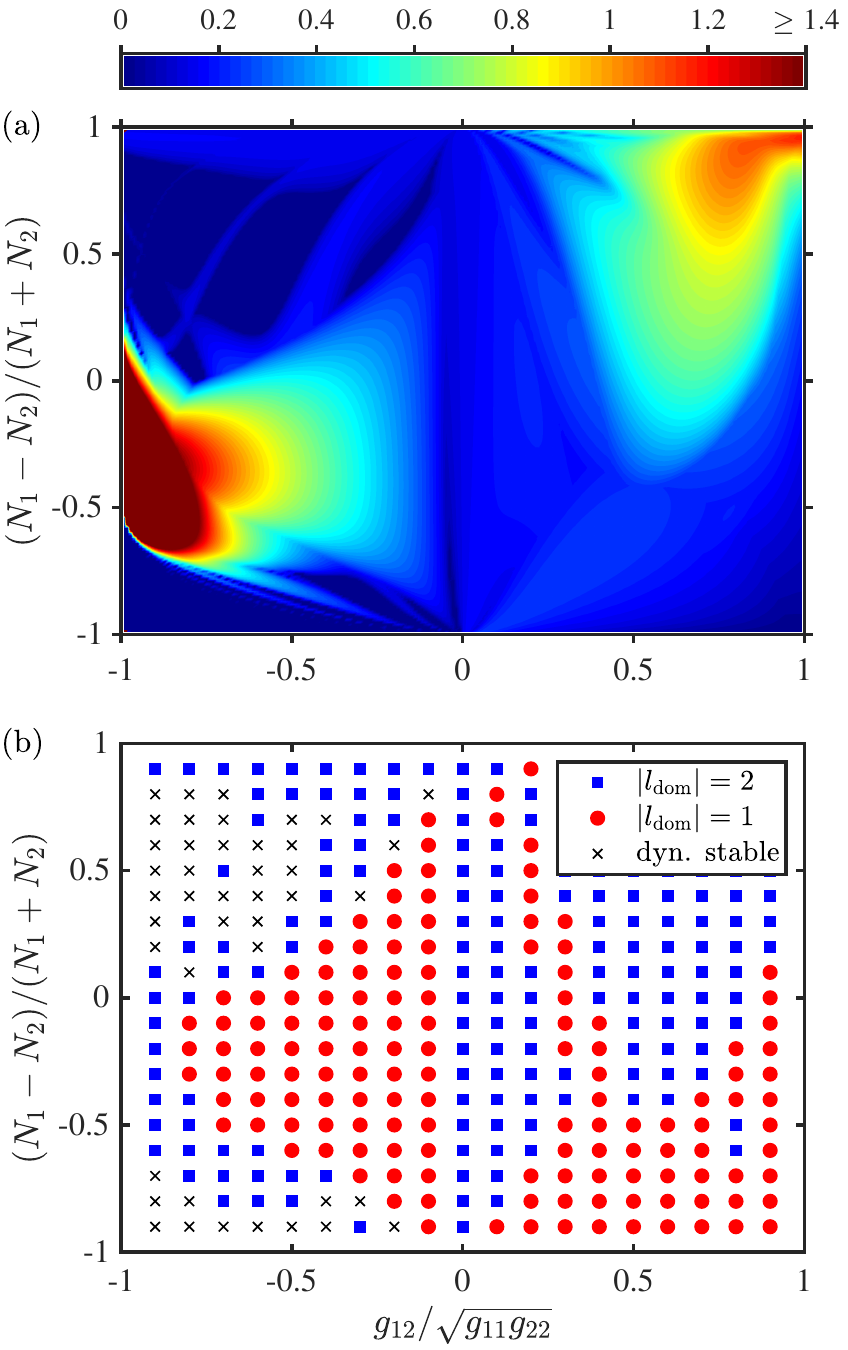}
\caption{\stabilitydiagramcaption{2}{-1}{\dynstabledescription}{}}\label{fig:maxim_2_-1}
\end{figure}

The stability diagrams of the $\tcv{2}{-1}$ vortex are presented in Fig.~\ref{fig:maxim_2_-1}. As in the case of the other counter-rotating vortex states (Figs.~\ref{fig:maxim_2_-2} and~\ref{fig:maxim_1_-1}), the region $\interparam < 0$ corresponds primarily to $\abs{\maxqpwinding}=1$, with small stability regions appearing close to the single-component limits $\polarization = \pm 1$. The region $\interparam > 0$ does not show dynamical stability. 

\begin{figure}[t]
\includegraphics[width=1.0\columnwidth,keepaspectratio]{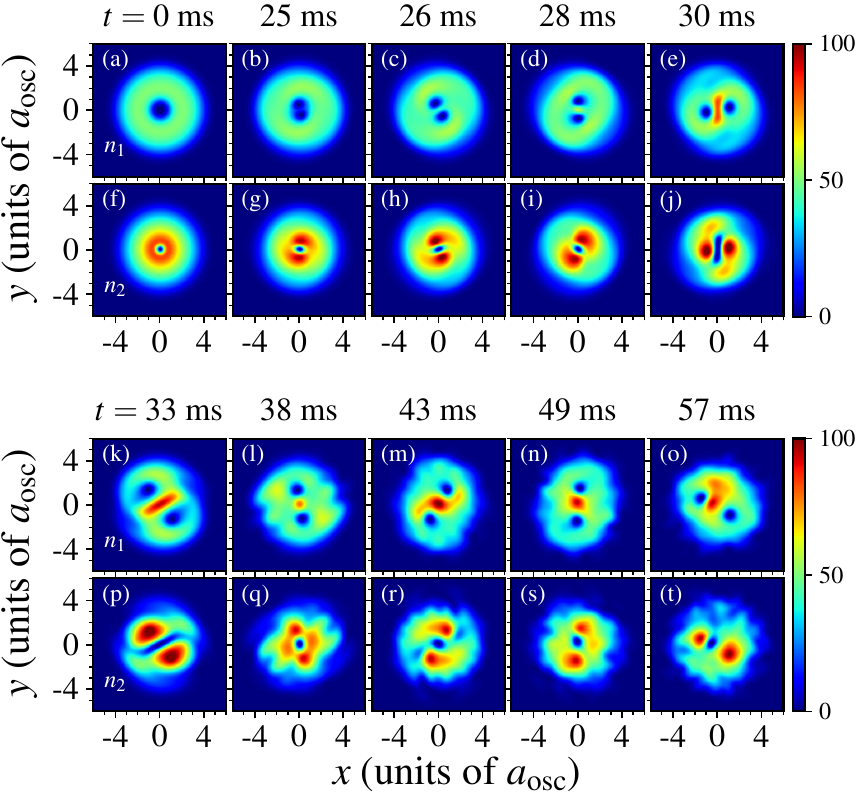}
\caption{\timeseriescaption{2}{-1}{0.7}{2}{0.437}{$n_1$}{$n_2$}{(a)--(e) and~(k)--(o)}{(f)--(j) and~(p)--(t)}}\label{fig:dyn_2_-1_a12_69}
\end{figure}

In Fig.~\ref{fig:dyn_2_-1_a12_69}, we illustrate the decay of the $\tcv{2}{-1}$ vortex with the case $\interparam = \num{0.7}$, which corresponds to $a_{12} = \num{69}\,\bohrradius$. For this state, the excitation with the largest imaginary part ($\num{0.437}\,\omega_1$) is obtained for $\abs{\qpwinding\mkern+1mu} = 2$. This leads to decay dynamics in which the doubly charged vortex in component~1 splits into two singly charged coreless vortices, as shown in Fig.~\ref{fig:dyn_2_-1_a12_69}(b). These singly charged vortices then orbit the trap center in a counterclockwise direction. During this splitting process, the singly charged antivortex in component~2 remains at the center until $t \approx \SI[mode=math]{43}{\milli\second}$. Around $t\approx \SI[mode=math]{30}{\milli\second}$, the core of this vortex becomes strongly elongated [Figs.~\ref{fig:dyn_2_-1_a12_69}(j) and~\ref{fig:dyn_2_-1_a12_69}(p)]; the phase field $\mathrm{arg}\left(\Psi_2\right)$ (not shown) becomes similar to that of a dark soliton, exhibiting a phase jump of $\pi$ across the elongated core while being approximately uniform within each of the two peaks in $n_2$. In this sense, the structure is reminiscent of the solitonic vortices occurring in elongated single-component BECs~\cite{Bra2002.PRA65.043612,Kom2003.PRA68.043617,Bec2013.NJP15.113028,Don2014.PRL113.065302} and superfluid Fermi gases~\cite{Ku2014.PRL113.065301}.

\subsection{\texorpdfstring{{\boldmath{$\tcv{2}{2}$}}}{(2,2)} vortex}

\begin{figure}[t]
\includegraphics[width=0.91\columnwidth,keepaspectratio]{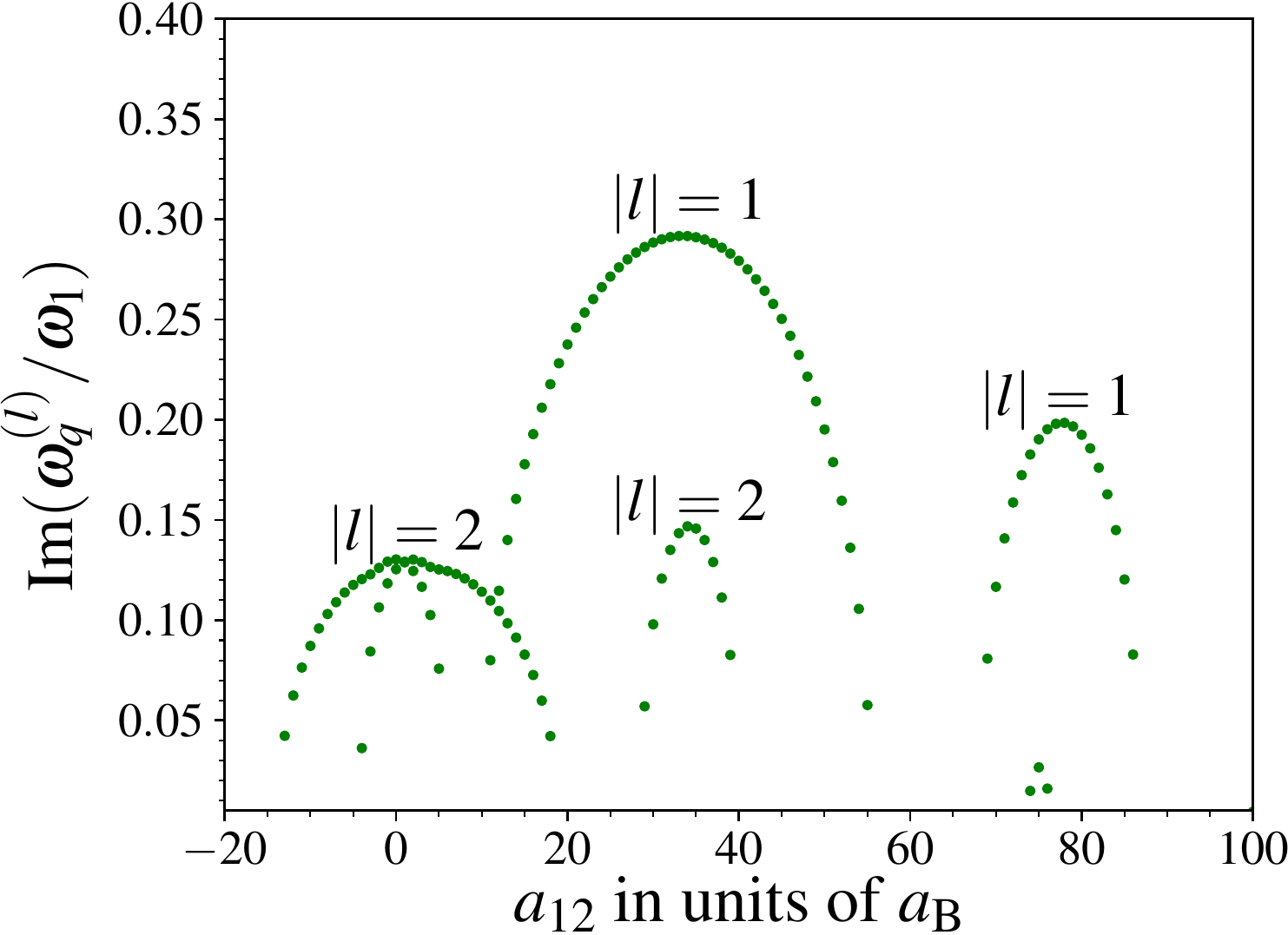}
\caption{\imaginarypartscaption{2}{2}}\label{fig:im_mode_evo_2_2}
\end{figure}

\begin{figure}[b]
\includegraphics[width=1.0\columnwidth,keepaspectratio]{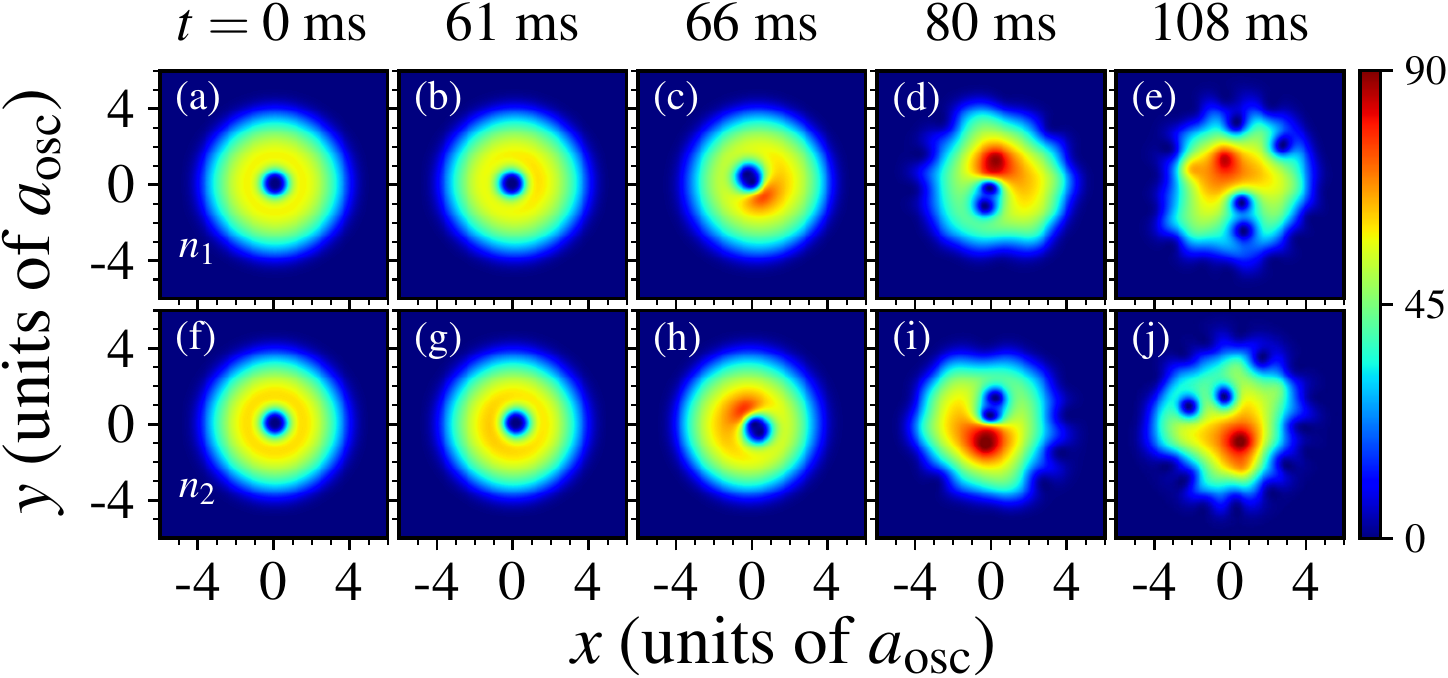}
\caption{\timeseriescaption{2}{2}{0.5}{1}{0.207}{$n_1$}{$n_2$}{(a)--(e)}{(f)--(j)}}\label{fig:dyn_2_2_a12_48.9}
\end{figure}

\begin{figure}[t]
\includegraphics[width=1.0\columnwidth,keepaspectratio]{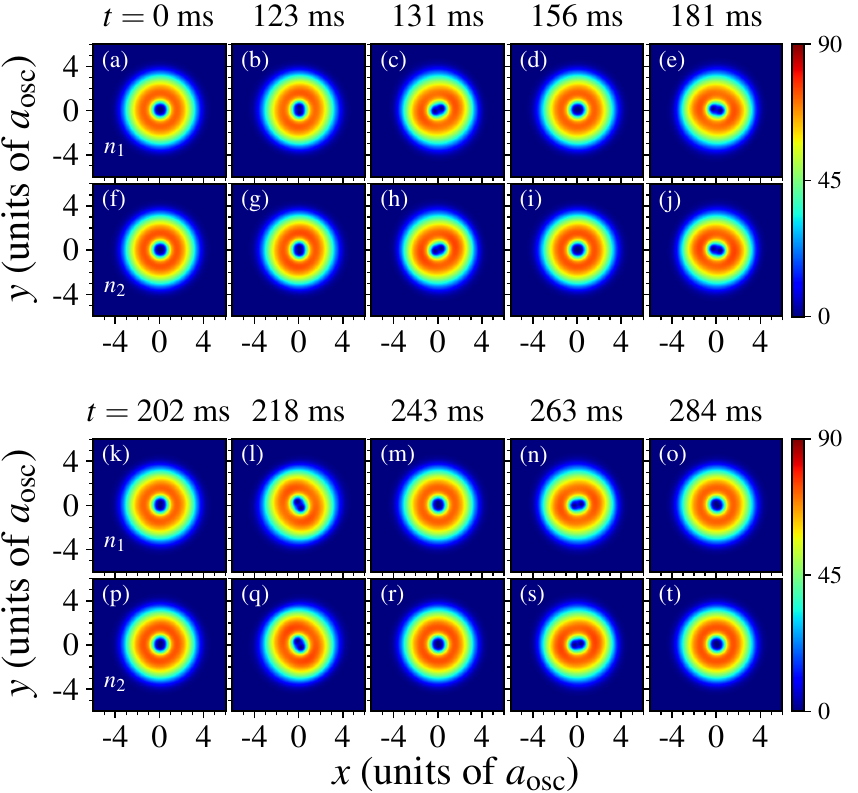}
\caption{\timeseriescaption{2}{2}{-0.1}{2}{0.100}{$n_1$}{$n_2$}{(a)--(e) and~(k)--(o)}{(f)--(j) and~(p)--(t)}}\label{fig:dyn_2_2_a12_-9.8}
\end{figure}

Figure~\ref{fig:im_mode_evo_2_2} shows the positive imaginary parts of the excitation frequencies as a function of $a_{12}$ for the $\tcv{2}{2}$ vortex in set~$\parameterset{2}$. For $\interparam\gtrsim \num{0.1}$, $\abs{\maxqpwinding} = 1$, whereas $\abs{\maxqpwinding} = 2$ for $\interparam\lesssim \num{0.1}$ (i.e., for $a_{12} \lesssim \num{10}\,\bohrradius$). There are also narrow windows of dynamical stability for $\num{0.6}\lesssim\interparam\lesssim \num{0.7}$ and $\num{0.9} \lesssim\interparam\lesssim \num{1}$.

To illustrate the dynamics associated with $\abs{\maxqpwinding}=1$, Fig.~\ref{fig:dyn_2_2_a12_48.9} shows the decay of the $\tcv{2}{2}$ vortex for $\interparam = \num{0.5}$, i.e., $a_{12} = \num{48.9}\,\bohrradius$. At $t \approx \SI[mode=math]{61}{\milli\second}$, the two doubly quantized vortices move to opposite sides of the trap center while orbiting it counterclockwise. Then, at $t\approx \SI[mode=math]{80}{\milli\second}$, each of them splits into two singly quantized vortices that continue the counterclockwise motion around the trap. In the composite-vortex notation, we can describe these events by the reaction sequence $\tcv{2}{2} \to \tcv{2}{0} + \tcv{0}{2} \to 2\times \tcv{1}{0} + 2\times \tcv{0}{1}$. The splitting process also induces significant motion of the center of mass of each condensate component, which leads at later times to the nucleation of additional vortices at the peripheries of the condensates [Figs.~\ref{fig:dyn_2_2_a12_48.9}(e) and~\ref{fig:dyn_2_2_a12_48.9}(j)].

The regime corresponding to $\abs{\maxqpwinding}=2$ is illustrated by the case $\interparam = \num{-0.1}$, i.e., $a_{12} = \num{-9.8}\,\bohrradius$. The associated time evolution is presented in Fig.~\ref{fig:dyn_2_2_a12_-9.8}. In contrast to the case $\abs{\maxqpwinding}=1$, both two-quantum vortices remain at the trap center until they each split at $t \approx \SI[mode=math]{130}{\milli\second}$. Subsequently, the resulting single-quantum vortices merge back together [Figs.~\ref{fig:dyn_2_2_a12_-9.8}(d) and~\ref{fig:dyn_2_2_a12_-9.8}(i)], and the system begins to exhibit split-and-revival behavior in resemblance to that shown in Fig.~\ref{fig:dyn_2_0_a12_69}.

\subsection{\texorpdfstring{{\boldmath{$\tcv{2}{-2}$}}}{(2,-2)} vortex}

\begin{figure}[t]
\vspace{0.00771331788cm}
\includegraphics[width=0.91\columnwidth,keepaspectratio]{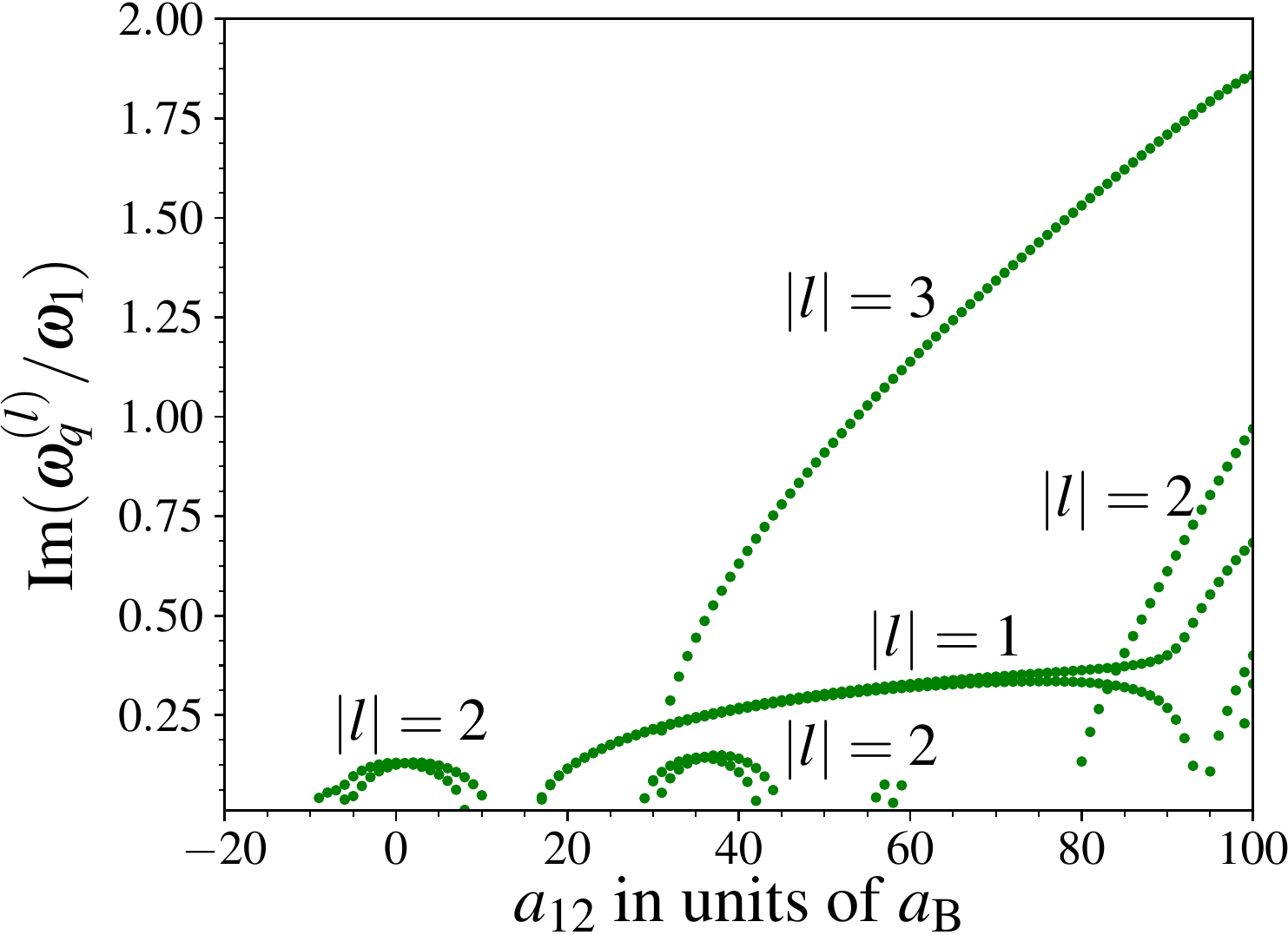}
\caption{\imaginarypartscaption{2}{-2}}\label{fig:im_mode_evo_2_-2}
\end{figure}

Figure~\ref{fig:im_mode_evo_2_-2} displays the positive imaginary parts of the excitation frequencies as a function of $a_{12}$ for the $\tcv{2}{-2}$ vortex in set~$\parameterset{2}$. As discussed in conjunction with Fig.~\ref{fig:maxim_2_-2}, for sufficiently strong intercomponent repulsion the dominant contribution to the dynamical instability comes from excitations with $\abs{\qpwinding\mkern+1mu}=3$; they produce the threefold splitting pattern illustrated in Fig.~\ref{fig:dyn_2_-2_a12_78.3}. Importantly, in view of the experimental verifiability of the threefold splitting, Fig.~\ref{fig:im_mode_evo_2_-2} reveals the $\abs{\qpwinding\mkern+1mu}=3$ peak to be much taller than the coexisting $\abs{\qpwinding\mkern+1mu}=1$ and $2$ peaks.
\bibliographystyle{apsrev4-1}
\bibliography{pak_bib_bec,tc-gv-excitations-manu}
\end{document}